\documentclass[10pt, a4paper]{article}
\usepackage{jheppub}
\usepackage{amsmath ,amsthm ,amssymb}
\usepackage{graphicx}
\usepackage{bbm}
\usepackage[utf8]{inputenc}
\usepackage{stmaryrd} 
\usepackage{tikz}

\theoremstyle{plain}
\newtheorem{thm}{Theorem}[section] 

\theoremstyle{definition}
\newtheorem{defn}[thm]{Definition} 
\newtheorem{exmp}[thm]{Example} 
\newtheorem{propos}[thm]{Proposition}

\title{Deformed graded Poisson structures, Generalized Geometry and Supergravity}
\author[a]{Eugenia Boffo,} \author[a]{Peter Schupp}
\affiliation[a]{Jacobs University Bremen,\\ Campus Ring 1, 28759 Bremen, Germany} 
\emailAdd{e.boffo@jacobs-university.de} 
\emailAdd{p.schupp@jacobs-university.de} 
\abstract{
In recent years, a close connection between supergravity, string effective actions and generalized geometry has been discovered that typically involves a doubling of geometric structures. We investigate this relation from the point of view of 
graded geometry, introducing an approach based on deformations of graded Poisson structures and derive the corresponding gravity actions. We consider in particular natural deformations of the $2$-graded symplectic manifold $T^{*}[2]T[1]M$ that are based on a metric $g$, a closed Neveu-Schwarz $3$-form $H$ (locally expressed in terms of a Kalb-Ramond 2-form $B$) and a scalar dilaton $\phi$.
The derived bracket formalism relates this structure to the generalized differential geometry of a Courant algebroid, which has the appropriate 
stringy symmetries, and yields a connection with non-trivial curvature and torsion on the generalized ``doubled" tangent bundle $E \cong TM \oplus T^{*}M$. Projecting onto $TM$ with the help of a natural non-isotropic splitting of $E$, we obtain a connection and curvature invariants that reproduce the NS-NS sector of supergravity in 10~dimensions. Further results include a fully generalized Dorfman bracket, a generalized Lie bracket and new formulas for torsion and curvature tensors associated to generalized tangent bundles. A byproduct is a unique Koszul-type formula for the torsionful connection naturally associated to a non-symmetric metric, which resolves ambiguity problems and inconsistencies of traditional approaches to non-symmetric gravity theories.}

\begin{document} 
\maketitle

\flushbottom

\section{Introduction}

Deformations of Poisson structures in classical physics and deformations of canonical commutation relations in quantum mechanics can be used to describe interactions. This approach is well-established in the context of electromagnetism and is an arguably slightly more general alternative to gauge theories. It allows for instance the inclusion of magnetic monopole sources \cite{Jackiw:1984rd}  (see also~\cite{Hanson:1974qy}) and it can also deal rather elegantly with first order actions. In the electromagnetic case, the deformation 
\[
\{p_\mu, p_\nu\} = e F_{\mu\nu}(x)  , \quad \{x^\mu,p_\nu\} = \delta^\mu_\nu , \quad \{x^\mu,x^\nu\} = 0
\]
of the canonical Poisson structure
is based on a local change of phase space coordinates $(x^\mu, p_\mu) \mapsto  (x^\mu, p_\mu + e A_\mu(x))$ generated by a gauge field $A_\mu(x)$. This is a simple application of Moser's lemma \cite{Moser1965}. Gauge transformations $\delta A_\mu = \partial_\mu \lambda$ correspond to canonical transformations of the deformed Poisson structure. The undeformed ``free'' Hamiltonian $H = p^2 /2m$ in conjuction with the deformed Poisson structure yields the correct Lorentz force $\dot p^\mu = e F^{\mu\nu} \dot x_\nu$ with $\dot x_\nu = p_\nu /m$. Globally, the deformation is non-trivial whenever the field strength belongs to a non-trivial cohomology class. 

 So far little is known about gravitational interactions from this point of view. Here we will show how to implement the approach in the context of supergravity. The deformation data that will enable us to formulate a Supergravity action is a Riemannian metric $g$, a closed  Neveu-Schwarz 3-form field strength $H$ and a dilaton scalar field $\phi$. 
Locally, on a contractible patch, the Neveu-Schwarz field can be expressed as  $H = dB$ in terms of the Kalb-Ramond 2-form $B$, which is defined up to 1-form $\Lambda$ (gauge) transformations $B \mapsto B + d\Lambda$. The 1-form gauge parameter $\Lambda$ is itself defined only up to a total derivative. Globally, this is the structure of an abelian bundle gerbe (a higher geometric analog of a line bundle). In analogy to the electromagnetic case mentioned above, the fields $g$, $B$ and $\phi$ generate a deformation via a local change of phase space coordinates. But unlike the electromagnetic case, the deformed Poisson structure cannot entirely be formulated in terms of gauge invariant quantities, since $\Lambda$ transformations act non-trivially -- as is in fact to be expected in view of the gerbe structure. The deformations are thus necessarily local, but they nevertheless extend to a globally well-defined structure. 

The appropriate geometric setting needed to accomodate all fields, turns out to be graded Poisson structures and it is closely related to Generalized Geometry.  The deformation data in fact defines a so-called generalized metric, but  this observation shall not be the main focus of this paper. 
Generalized Geometry unifies aspects of Riemannian, symplectic and complex geometry. It typically involves the study of a doubled (tangent plus cotangent) bundle $TM \oplus T^{*}M$ with structure group $O(d,d)$, or extensions of it, interpreted as a generalized tangent bundle. 
The natural notion of  symmetry on the generalized tangent bundle is encoded in Courant algebroids with an ad-invariant pairing on sections, an anchor map into  tangent space and a Dorfman bracket. Unlike the more familiar Lie-bracket, the Dorfman bracket is not anti-symmetric in order to assure suitable integrability properties and it satisfies a restricted Jacobi identity.  Compatibility conditions between these objects must also hold. Exact Courant algebroids $E \cong TM \oplus T^{*}M$ have been classified by  \v{S}evera  \cite{Severa:2017oew} in terms of the third cohomology class $H^{3}(M, \mathbb{R})$.

In the recent past the generalized differential geometry of $TM \oplus T^{*}M$ has been exploited to show -- with various suitable assumptions -- that supergravity, as the supersymmetric theory of gravity in its own right, but also as the effective field theory of superstrings of type IIA and IIB, can be described as some kind of Einstein's General Relativity on  this doubled vector bundle. For example, the works \cite{Coimbra:2011nw}, \cite{Aldazabal:2013mya}, \cite{Jurco:2016emw} and others show this in the framework of Generalized Geometry. In Double Field Theory, where also the coordinates of the base manifold (spacetime) are doubled, similar results -- upon suitable projection onto standard target spacetime -- were found, see e.g.\ \cite{Hohm:2010jy}.

As mentioned above, here we shall take a different approach based on deformations of graded Poisson structures.  It is known that the aforementioned Generalized Geometry structures are special cases of graded Poisson algebras in the derived bracket formalism: The $2$-graded symplectic manifold $T^{*}[2]T[1]M$, admitting a Hamiltonian (shifted) vector field, with its sheaf of graded Poisson algebras generated by the polynomial functions, was related to the exact Courant algebroid on $T[1]M \oplus T^{*}[1]M \simeq T^{*}M \oplus TM$ in \v{S}evera's letter~$7$ \cite{Severa:2017oew} to Alan Weinstein.  Roytenberg, in \cite{Roytenberg:2002nu} and in his PhD thesis \cite{math/9910078}, further analysed graded symplectic manifolds and found a 1-1 correspondence between symplectic $NQ$-manifolds of degree 2 and Courant algebroids.
The key was to notice that the bracket and the pairing are derived brackets of the Poisson bracket with the Hamiltonian as differential. The relation between Courant algebroid and derived brackets and other applications to Lie algebroids are explained by Kosmann-Schwarzbach in \cite{KosmannSchwarzbach:2003en}.
Graded Poisson algebras and graded Lie algebras are also relevant in the context of the BRST and BV quantization of the path integral of field theories with local symmetries. 
Another closely related and fruitful setup for the exploitation of the rich structures of graded manifolds are the AKSZ models \cite{Alexandrov:1995kv}. They associate topological field theories to graded symplectic manifolds by lifting the graded and symplectic construction on a pair of manifolds $M, N$ to the mapping space $\text{Map}(M,N)$. For further details see e.g.\ the review on supergeometry \cite{Cattaneo:2010re}. Worth mentioning, amongst the numerous works, is the exploration of the AKSZ construction by Ikeda \cite{Ikeda:1993fh}, Schaller and Strobl in \cite{Schaller:1994es}, and Cattaneo and Felder in \cite{Cattaneo:2001ys} and \cite{Cattaneo:2001bp} for $N = T[1]N_{0}$ related to the Poisson $\sigma$-model.

It is natural to conjecture that the ``generalized General Relativity nature'' of supergravity can be described in terms of the differential graded symplectic manifold $T^{*}[2]T[1]M$, with $M$ being $d$-dimensional target spacetime. Indeed, this is the program of this paper: Starting from a deformation of the graded Poisson algebra encoded in local vielbeins, we reconstruct the NS-NS sector of $10$-dimensional supergravity from a derived generalized connection and corresponding curvature invariant. The construction uses the derived bracket approach to Generalized Geometry and some additional natural geometric structures, which we shall describe in detail.

\paragraph{Structure of the article:} In the following section \ref{very_first} we will 
review the necessary background on Generalized Geometry and graded symplectic manifolds. The notation will be fixed in this part. In section \ref{first} the general deformed, i.e.\ non-canonical Poisson structure on $C^{\infty}(T^{*}[2]T[1]M)$ is analysed and its Courant algebroid counterpart on  on $TM \oplus T^{*}M$ is computed via the derived bracket approach. Section \ref{middle} is devoted to a survey of differential geometry in this setting and in particular to generalized connections and associated tensors. 
This section contains several new results. Definitions for a generalized Lie bracket, torsion and curvature tensors, connection symbols of the first kind are given and a general proposition \ref{propos} that relates these objects to the generalized Dorfman bracket. We will see how a connection with non-trivial curvature arises from the derived bracket formalism. 
The section includes a comparison with previously proposed notions of torsion in the context of Generalized Geometry. 
Finally, in section \ref{body}, all these methods are applied with a concrete deformation based on a Riemannian metric $g$, a $2$-form $B$ and a scalar (dilaton) $\phi(x)$, eventually yielding the effective Lagrangian for type II closed strings. An interesting side-result is a generalization of the  Koszul formula to a non-symmetric metric setting with dilaton.

\section{Summary of essential notions}
\label{very_first}

The original works  on Generalized Geometry, which received a lot of attention in the early $2000$'s are usually considered those by Hitchin \cite{Hitchin:2010qz} and by his student Gualtieri \cite{Gualtieri:2003dx}, to which we refer for a nice explanation of the underlying concepts of Generalized Geometry. The key underlying structure is that of a Courant algebroid:
\begin{defn}
A \emph{Courant algebroid}  $(TM \oplus T^{*}M \equiv E, \rho,  [\cdot, \cdot], \langle \cdot, \cdot \rangle)$ consisting of a vector bundle with a point-wise non-degenerate symmetric bilinear form on sections $\langle \cdot, \cdot \rangle$, together with a bracket $[ \cdot, \cdot ] : \Gamma(E) \times \Gamma(E) \rightarrow \Gamma(E)$ and a map $\rho : E \rightarrow TM$ (called the \emph{anchor}), such that:
\begin{enumerate}
\item $[e, [e^{\prime}, e'']] = [[e, e'],e''] + [e',[e,e'']]$\, , \, \, $\forall \, e, e', e'' \in \Gamma(E)$,
\item $\langle e, [e',e'] \rangle = \frac{1}{2}\rho(e) \langle e',e' \rangle$,
\item $\langle [e,e'], e' \rangle = \frac{1}{2} \rho(e) \langle e', e' \rangle$ .
\end{enumerate}
This is a minimal set of axioms; other possible relations can be shown to follow from these three conditions. For example the anchor map can be proven to be  a homomorphism of brackets, $\rho( [e, e']) = [\rho(e), \rho(e') ] \vert_{TM}$ and the last two axioms can be polarized (using $e^{\prime} := \tilde{e} + \hat{e}$):
\begin{align}
\langle e, [\tilde e,\hat e] +  [\hat e,\tilde e] \rangle &= \rho(e) \langle \tilde e, \hat e \rangle \,, \\
\langle [e,\tilde e], \hat e \rangle + \langle \tilde e, [e,\hat e] \rangle &=  \rho(e) \langle \tilde e, \hat e \rangle \,.
\end{align}
In this form the last axiom also yields the Leibniz rule of $[ \cdot, \cdot]$ for the second entry only:
\begin{equation}
[e, fe^{\prime}] = \left( \rho(e) f \right) e^{\prime} + f [e, e^{\prime}], \quad \forall f \in C^{\infty}(M). \label{CLeibniz}
\end{equation}
A \emph{Lie algebroid}  $(E, \rho, [\cdot, \cdot])$ fulfills the first axiom and the Leibniz rule and its bracket is skew-symmetric.
\label{Courant} 
\end{defn}

\noindent
Courant algebroids for which the following short exact sequence holds
(i.e.\ if $j$ is an injective inclusion map, $\rho$ is surjective and $\text{Im} j = \text{Ker} \rho$)
\begin{equation}
0 \rightarrow T^{*}M \xrightarrow{j} E \xrightarrow{\rho} TM \rightarrow 0 \, ,
\label{seq}
\end{equation}
are called exact.  A splitting of the exact Courant algebroid is an embedding $s: TM \rightarrow E$ such that $\rho \circ s = \text{id}$. It is then straightforward to show that $s \oplus j: TM \oplus T^*M \rightarrow E$ is a vector space isomorphism. The splitting $s$ is not unique and it does not need to be isotropic -- in the sense that the bilinear pairing $\langle \cdot, \cdot \rangle$ vanishes on its image -- in order to establish the isomorphism. We shall in fact later use a non-isotropic splitting to project a generalized connection to the tangent bundle. 

The canonical example of an exact Courant algebroid is given by the  bundle $E = TM \oplus T^{*}M \simeq T^{*}[1]M \oplus T[1]M$ that is the Whitney sum of the tangent space and the cotangent space together with their natural pairing and with $\rho$ being the natural projection onto $TM$.
The non-skew-symmetric bracket for the Courant algebroid most commonly used in Generalized Geometry is the Dorfman bracket.
(Its antisymmetrization, called the Courant bracket, will not be used here, as the axioms would be slightly different -- the Jacobi identity in the definition would hold only up to a  Jacobiator.)
\begin{defn}[Dorfman bracket]
Let $U = X + \eta$, $V = Y + \zeta$ denote a pair of local sections of $T M\oplus T^*M$, i.e. $X,Y \in \Gamma(T M)$ and $\eta, \zeta \in \Gamma(T^{*}M)$. Then the \emph{Dorfman bracket} $[U, V]_{\text{Dorf}}$ is given by:
\begin{equation}
[U, V]_{\text{Dorf}} := [X,Y] + \mathcal{L}_{X} \zeta - \iota_{Y} d \eta \, ,
\label{DDD}
\end{equation}
where the bracket on vector fields is the standard Lie bracket.\\
The Dorfman bracket can be twisted by a 3-form $H \in \Omega^3 (M)$:
\begin{equation}
[U,V]_H = [U, V]_{\text{Dorf}} + \iota_X \iota_Y H \, . \label{twistedDDD}
\end{equation}
\end{defn}

Exact Courant algebroids have been classified by \v{S}evera \cite{Severa:2017oew} in terms of the third cohomology class $H^{3}(M, \mathbb{R})$. Let us briefly recall the construction without going into details:  A natural choice for the inclusion map $j$ is the adjoint  $\rho^*$ of $\rho$ that is defined with respect to the bilinear form of the Courant algebroid and the natural pairing of $TM$ and $T^*M$. The inclusion $\rho^*$ is isotropic in the sense that the bilinear form $\langle \cdot, \cdot \rangle$ vanishes on its image. It can be shown that one can also always find an isotropic splitting $s_{\text{iso}}$. Via $s_{\text{iso}}\oplus \rho^*$ the exact Courant algebroid is then seen to be isomorphic to the aforementioned canonical example with the twisted bracket (\ref{twistedDDD}). Any two isotropic splittings are related by a $B$-transform that maps $H \mapsto H +dB$. Hence, exact Courant algebroids are classified by $[H] \in H^{3}(M, \mathbb{R})$. We shall later extract a generalized connection from a deformed Courant algebroid. In view of \v{S}evera's classification this will clearly require some additional geometric structure like a particular choice of splitting or frame, which however turns out to be quite natural. We will comment on this in due course.

Graded symplectic manifolds are manifolds very rich in structure and properties.  For the basics on graded manifolds and algebras, we refer to \cite{Cattaneo06gradedpoisson}. We will focus on $T^{*}[2]T[1]M$, the shifted cotangent bundle of the shifted tangent bundle of a manifold $M$, which we shall interpret as (target) spacetime. Darboux's theorem still holds despite the shifts in the grading. In an affine Darboux chart corresponding to a chart $\{x^i\}$ on $M$ (which also coincides with a local trivialization of the bundle), coordinates on $T^{*}[2]T[1]M$ are given by $(x^{i}, \xi_{\alpha}, p_{i})$, where $\left(\xi_{\alpha} \right):= \left( \chi_{i}, \theta^{i} \right)$ are $2d$ odd coordinates on the fibers of $T^*[1] \oplus T[1] M$,  $(x^{i})$ are coordinates  on the base $M$ and the momenta $(p_{i})$ are coordinates  on the fibers of $T^*[2]M$. Weights $\lvert \cdot \rvert \in \mathbb{Z}$ are assigned as $\lvert x^{i} \rvert =0, \, \lvert \xi_{\alpha} \rvert=1, \, \lvert p_{i} \rvert = 2$. 
From the graded geometry point of view the choice of $\xi_\alpha$ (i.e. $\chi_i$ and $\theta^i$) is part of the local Darboux coordinate choice, from a bundle point of view it is a choice of preferred frame (coordinate basis and its dual) and we can identify $(\chi_i, \theta^i)$ with $(\partial_i, dx^i)$. \emph{All coordinate-dependent expressions that follow will refer to the local canonical graded coordinate system that we have described here.} 

The canonical symplectic form of $T^{*}[2]T[1]M$ has degree $2$ and is given by
\begin{equation} \omega = dx^{i} \wedge dp_{i} + d\chi_{i} \wedge d\theta^{i} \,,
\label{symplectic}
\end{equation}
while
the corresponding canonical Poisson brackets have degree $-2$:
\begin{equation}
\begin{array}{l}
\{p_{i}, x^{j} \} = \delta_{i}^{\; j}, \quad  \{p_{i} , \xi_{\alpha} \} =0 , \quad
\{p_{i}, p_{j} \} = 0   \,,\\
\{\chi_i,\theta^j\} = \delta_i^j, \quad \{\chi_i,\chi_j\} = 0, \quad \{\theta^i,\theta^j\} = 0\, .
\label{Pois_br_def_canonical}
\end{array}
\end{equation}
The last line can be written more compactly as $\{\xi_{\alpha}, \xi_{\beta} \} = \eta_{\alpha \beta}$ with the $O(d,d)$-invariant pairing $\eta$.

Before reviewing the derived brackets construction that leads to the fundamental result on the relation between this dg-symplectic manifold of degree $2$ and the Courant algebroid $(TM \oplus T^{*}M, \rho,  [ \cdot, \cdot], \langle \cdot, \cdot \rangle)$, we shortly recall the notion of a  homological vector field and under which conditions it gives rise to a Hamiltonian in this graded setup.
\begin{defn}
A \emph{homological vector field} is a vector field $Q$ belonging to the degree $+1$ shifted vector fields $\mathfrak{X}\Pi(E)$,  such that $\{Q,Q\}_{+} = 2 Q^{2} =0$.\\
When such a vector field preserves the symplectic form $\omega$, i.e. $\mathcal{L}_{Q} \omega =0$, it is a symplectic homological vector field.
$Q$ is symplectic homological iff it is \emph{Hamiltonian} ($\iota_{Q} \omega  = d \xi$, for $d \xi$ some exact form) and $n+1 \neq 0$, $n$ being the degree of $\omega$.\\
If $n \neq -1$, $Q$ is hence recovered from the Poisson bracket (of degree $-n$) with a corresponding Hamiltonian of weight $1+n$, denoted by $\Theta$: \
$Q = \{ \Theta, \cdot \}$.
\label{HOM_Q}
\end{defn}
A Hamiltonian $\Theta$ for $T^{*}[2]T[1]M$ must therefore be of degree $3$. The most general degree $3$ Hamiltonian $\Theta$ that can be written down consists of a kinetic term linear in the momenta and of a potential cubic in the degree-$1$ coordinates:
\begin{equation}
\Theta = \xi_{\alpha} \rho^{\alpha i}(x) p_{i} + \frac{1}{3!} C^{\alpha \beta \gamma}(x) \, \xi_{\alpha} \xi_{\beta} \xi_{\gamma} \, ,
\label{TH_roy}
\end{equation}
where $C^{\alpha \beta \gamma}(x) \, \xi_{\alpha} \xi_{\beta} \xi_{\gamma} =: C$, due to the odd parity of the $\xi$'s,  
is a completely antisymmetric function in the sheaf $\mathcal{O}_{3}$ (the set of polynomials of degree $3$). It can be interpreted very naturally as the tensor representing all the T-dual stringy geometrical and non-geometrical ``fluxes'': $R^{abc}$, $Q_{a}^{\; \; bc} $, $f_{ab}^{\; \; \; c} $ and $H_{abc}$. 
The position-dependent matrix $\rho^{\alpha i}(x) $ will play the role of the anchor map of a Courant algebroid.
The Hamiltonian $\Theta$ must satisfy the structure or master equation (corresponding to $Q^{2}=0$, as can be seen from the Jacobi identity and the definition \ref{HOM_Q}):
\begin{equation}
\{ \Theta, \Theta \}=0 \, .
\label{master}
\end{equation}
This implies various constraints on the rank-$3$ tensor $C$, on $\rho^{\alpha i}(x)$ and compatibility with the graded Poisson structure must also be checked.

The master equation \eqref{master} together with the graded Jacobi identity for the Poisson bracket implies the axioms of a Courant algebroid \ref{Courant} that can be obtained as a derived structure. The bracket, the anchor map and the pairing are defined by
\begin{align}
\{\{e_{1}, \Theta \}, e_{2} \} = & \, \{\{ \Theta, e_{1} \} , e_{2} \} = [e_{1}, e_{2}] \, , \label{derived_brackets} \\
\{\{ e, \Theta \}, f \} =& \, \{\{ \Theta, f\}, e\} = \rho(e) f \, , \label{der_br_func} \\
\{ e_{1}, e_{2} \} = & \, \langle e_{1}, e_{2} \rangle \, , \label{pair}
\end{align}
for $e, e_{1}, e_{2} \in \mathcal{O}_{1}$ being functions of degree 1 on $T^{*}[2]T[1]M$. These 
functions can be identified with elements of the space of sections $\Gamma(TM \oplus T^{*}M)$. From now on we will always identify the $\xi_\alpha$'s with a basis of sections, and all following coordinate expressions will refer to the local coordinate basis introduced in this section. Moreover the anchor $\rho \in \text{Hom}(E, TM)$ is understood as the same map appearing in \eqref{TH_roy} and the constraints due to the master equation impose some restrictions on it. 
The proof of the correspondence between the dg-symplectic manifold and the Courant algebroid can be given as follow. The first axiom is a consequence of the (graded) Jacobi identity for the Poisson graded algebra: 
\begin{align}
 [ \xi_{1}, [\xi_{2}, \xi_{3}] ] = \{ \{\Theta, \xi_{1} \}, \{ \{ \Theta, \xi_{2} \}, \xi_{3} \}\} &= \{ \{ \{ \{ \xi_{1}, \Theta \} , \xi_{2} \}, \Theta\}, \xi_{3}\} + \{ \{ \Theta, \xi_{2} \}, \{ \{ \Theta, \xi_{1} \}, \xi_{3} \} \} \notag \\
\; & \; \; \; \;+ \frac{1}{2} \{ \{ \{ \{\Theta, \Theta \}, \xi_{1} \}, \xi_{2} \}, \xi_{3} \} \notag \\
\; & = [ [ \xi_{1}, \xi_{2} ], \xi_{3}] + [ \xi_{2}, [ \xi_{1}, \xi_{3} ]] + \frac{1}{2} \{ \{ \{ \{\Theta, \Theta \}, \xi_{1} \} , \xi_{2} \}, \xi_{3} \}. \, \label{Jacobi}
\end{align}
The last term is zero due to the master equation \eqref{master} and the axiom is verified.\\
For what concerns the remaining two axioms, the (graded) Leibniz rule and the (graded) Jacobi identity give directly:
\begin{align}
 \rho(\xi_{1}) \left\langle \xi_{2}, \xi_{2} \right\rangle = \{ \{ \Theta, \xi_{1} \}, \{ \xi_{2}, \xi_{2} \} \} &= 2\{ \{ \{ \Theta, \xi_{1} \} , \xi_{2}  \}, \xi_{2} \} = 2 \left\langle \left[ \xi_{1}, \xi_{2}\right], \xi_{2} \right\rangle & \text{axiom} \; 2, \label{2_ax}\\
\; & = 2 \{ \xi_{1} , \{ \{ \Theta , \xi_{2} \}, \xi_{2} \}\} = 2 \left\langle \xi_{1}, \left[ \xi_{2}, \xi_{2} \right] \right\rangle & \text{axiom} \; 3, \label{3_ax}
\end{align}
Hence, we have shown that when a Hamiltonian $\Theta$ in \eqref{TH_roy} on the graded symplectic manifold $T^{*}[2]T[1]M$ is given, the data and the algebraic relations for $T^{*}[2]T[1]M$ imply those for a Courant algebroid on $TM \oplus T^{*}M$, as defined in \eqref{Courant}.

For instance, let us apply these considerations to the canonical case, whose symplectic form is given in \eqref{symplectic}. Equation \eqref{pair} immediately tells us that the pairing is the $O(d,d)$-invariant constant~$\eta$. Furthermore, the anchor $\rho^{\alpha i}(x) = \left(\rho^{a i} (x) , \rho_{a}{}^{ i}(x) \right)^{T}$ must satisfy the following differential and algebraic equations
\begin{equation}
\begin{cases}
\rho_{[\beta \vert}{}^{k}\partial_{j} \rho_{\vert \alpha]}{}^{i}(x) = 0, \\
\delta^{a}{}_{ b} \, \rho_{a}{}^{j} (x) \rho^{b i}(x) = 0,
\end{cases}
\label{anchor}
\end{equation}
as found from $\{ \Theta, \Theta \}= 0$. These relations can be solved by choosing $\rho \in \text{Hom}(E, TM)$ to be constant and null on forms, i.e. it is the projector onto the tangent space. Finally, the derived bracket for the Courant algebroid \eqref{derived_brackets} is precisely the Dorfman bracket given in equation \eqref{DDD}, possibly twisted with the $H$ $3$-form introduced previously as the component of $C$ that lives only in $T^{*}M$. According to \v{S}evera and Weinstein classification of Courant algebroids, $H$ is the representative in the third cohomology class $H^{3}(M, \mathbb{R})$.

Hence from the canonical graded symplectic structure of $T^{*}[2]T[1]M$ one can reconstruct the canonical example of a Courant algebroid, $(TM \oplus T^{*}M, \rho, [\cdot, \cdot ]_{\text{Dorf}}, \eta )$.

\section{Graded Poisson algebra of $T^{*}[2]T[1]M$}
\label{first}
\subsection{Deformed graded Poisson brackets}
\label{Defo_Pois}
Consider again the graded symplectic manifold $(T^{*}[2]T[1]M, \omega)$ just outlined. The most general deformation of the canonical Poisson structure (\ref{Pois_br_def_canonical}), which we require to preserve the $x$-$p$ bracket and (of course) still obeys the graded Jacobi identity and the graded Leibniz rule is given by the following brackets \eqref{Pois_br_def}. They are listed in the left column in components and in the right column in a coordinate free notation, where $C^{\infty}(T^{*}[2]M) \ni w,v :=  v^{i}(x) p_{i}$, $C^{\infty}(T^{*}[1]M \oplus T[1]M) \ni V, U :=  U^{\alpha}(x) \xi_{\alpha}$ and $f(x) \in C^{\infty}(M)$.
\begin{equation}
\begin{array}{ll}
\{p_{i}, x^{j} \} = \delta_{i}^{\; j}, & \; \;  \; \{v, f \} = v(f) , \\
\{\xi_{\alpha}, \xi_{\beta} \} = G_{\alpha \beta}(x) , & \; \; \; \{ U, V \} =G(U,V) ,\\
\{p_{i} , \xi_{\alpha} \} =\mathit{\Gamma}^{\beta}_{\; \; i \alpha}(x) \, \xi_{\beta} , & \; \; \; \{v, U \} = \nabla_{v} U, \\
\{p_{i}, p_{j} \} = \text{R}^{\alpha \gamma}_{\; \; \; \; ij}(x) \, \xi_{\alpha} \xi_{\gamma}, & \; \;  \; \{v, w\} = [v,w]_{\text{Lie}} + \text{R}(v,w)  .
\label{Pois_br_def}
\end{array}
\end{equation}
This is a non-degenerate Poisson structure if some restrictions are placed on $G$, $\nabla$ and $\text{R}$: $G$ should be a $2d \times 2d$ invertible symmetric block matrix, $G_{\alpha \beta} \in S^{2}(T^{*}[1]M \oplus T[1]M)$, as the identities of the Poisson algebra, the degree counting and the anticommutativity of the bracket imply directly. It should be considered as a generalization of the canonical $O(d,d)$-invariant pairing between vector fields and forms 
in a sense that will be clear from the particular deformation one wants to analyze.

Similarly, from the bracket between the functions linear in the momenta and those linear in~$\xi$ we must expect a degree $1$ object with the properties of a connection. The Jacobi identity with another degree $1$ object implies that this connection $\nabla$ is metric with respect to $G_{\alpha \beta}(x)$. In fact, according to Roytenberg's work \cite{Roytenberg:2002nu}, the action of the polynomial functions of degree $2$ (which are just those linear in $p$, because those quadratic in $\xi$ actually act trivially) on the functions of lower degree comes from a Lie algebroid action preserving $G$. In formal terms, the algebra of sections of the Lie algebroid is the Atiyah algebra of \emph{covariant differential operators} on $T^{*}[1]M \oplus T[1]M$ preserving the metric. The connection $\nabla$ can be understood as a splitting of the Atiyah sequence associated to symplectic $N$-manifolds of degree 2 \cite{Roytenberg:2002nu}.
The Poisson bracket of two momenta is given by the curvature $2$-form of the aforementioned connection, $\text{R}_{ij} \in \Lambda^{2} \left(T^{*}[2]M \right)$, as implied by the Jacobi identity with $\xi$, the commutativity of the bracket and the resulting degree ($2$).
The symplectic form $\omega$, which corresponds to the inverse of the Poisson bivector, is
\begin{align}
\omega = & \,  dx^{i} \left[ \text{R}^{\alpha \beta}_{\; \; \, ij} \xi_{\alpha} \xi_{\beta} + \xi_{\beta} \mathit{\Gamma}^{\beta}_{\; i \alpha} \left( G^{-1} \right)^{\alpha \gamma} \mathit{\Gamma}^{\delta}_{\; j \gamma} \xi_{\delta}\right]\wedge dx^{j} - dx^{i} \left[ \xi_{\gamma} \mathit{\Gamma}^{\gamma}_{\; i \alpha} \left( G^{-1} \right)^{\alpha \beta} \right] \wedge d\xi_{\beta} \notag \\
\, & + d\xi_{\alpha} \left[ \left(G^{-1} \right)^{\alpha \beta} \right] \wedge d\xi_{\beta}  + dx^{i} \wedge dp_{i} . \label{sy_def}
\end{align}
The master equation \eqref{master} places further restrictions on $G$, $\nabla$ and $\text{R}$ resulting in expressions that can be decomposed into a set of degree 4 equations, each involving a different monomial in the degree 1 and 2 generators.  In particular the degree $4$ equation for a quadruplet of $\xi$'s is
\[
\frac{2}{3!} \xi_{\alpha} \rho^{\alpha i} \left\{ p_{i}  , C^{\beta \gamma \delta }(x) \xi_{\beta} \xi_{\gamma} \xi_{\delta} \right\} + \frac{1}{3! 3!} C^{\alpha \beta \gamma}(x) C^{\delta \epsilon \varphi}(x) \left\{ \xi_{\alpha} \xi_{\beta} \xi_{\gamma}, \xi_{\delta} \xi_{\epsilon} \xi_{\varphi} \right\} 
+ \xi_{\alpha}  \xi_{\beta} \rho^{\alpha i}  \rho^{\beta l} \left\{ p_{i}, p_{l} \right\} = 0 ,
\]
which gives
\[
\frac{2}{3!} \nabla_{\rho(\xi)} C(x) + \frac{1}{3! 3!} C(x) \mathrel{\llcorner} C(x) = - \xi_{\alpha}  \xi_{\beta} \rho^{\alpha i}  \rho^{\beta l} \text{R}^{\gamma \tau}_{\; \; \; \; il}(x) \xi_{\gamma} \xi_{\tau} ,
\]
where $ \nabla_{\rho(\;)}$ is the differential graded connection on the generalized tangent bundle $T^{*}[1]M \oplus T[1]M$ induced via the anchor map $\rho$ from the ordinary affine connection $\nabla$,
 and the interior product is performed with $G$. This equation can be solved for $\text{R}$ as a function of the fluxes $C$, of the metric $G$ and of the anchor $\rho$. Therefore, without loss in generality, we can take $\text{R} = 0$ in \eqref{Pois_br_def} and focus on the fluxes to keep track of the modifications induced by the curvature. For a complete analysis on the non-geometric fluxes and their Bianchi identities in the canonical dg-symplectic $T^{*}[2]T[1]M$ one can consult \cite{Heller:2016abk}. From now on the Poisson bracket of the momenta and hence $\text{R} $ is set to zero unless stated otherwise. (Later in the paper we will consider a different ``derived'' connection with non-zero curvature.)
In the absence of curvature the connection $\nabla$ is thus forced to be the curvature-free but generically torsionful (in some wider sense) metric connection said to be of Weitzenb\"{o}ck type. Weitzenb\"{o}ck connections have already been used before in \cite{Penas:2018mnc} and \cite{Berman:2013uda} in the related context of Double Field Theory, and proven useful in the derivation of Supergravity theories. In order to avoid potential confusion regarding global implications, let us mention that the above discussion of absorbing $R$ in the fluxes was done in a local coordinate chart and in the present context, the Weitzenb\"{o}ck connection can also be traced back to fixing a local frame for the degree-$1$ coordinates, or equivalently a generalized local vielbein $\text{E}$. We are certainly \emph{not} requiring a global vielbein or parallelizable manifold. The metric $G$ too has an interpretation in term of $\text{E}$, as that one metric to which the $O(d,d)$ $\eta$ is reduced by the vielbein in a local neighborhood.

The master equation for the bracket between $\xi_{\alpha} \left(G^{-1} \right)^{\alpha \beta} \rho_{\beta}^{\; \; i} p_{i}$ and itself gives the remaining couple of relations. These involve the degree $4$ objects built from a pair of momenta and from a pair of $\xi$'s together with one $p$:
\[
\left\{ \xi_{\alpha} , \xi_{\beta} \right\} \left(G^{-1}\right)^{\alpha \gamma} \rho_{\gamma}^{\; \, i} \left(G^{-1}\right)^{\beta \delta} \rho_{\delta}^{ \; \, j} p_{i} p_{j} = 0 , \quad 2 \xi_{\alpha} \left(G^{-1}\right)^{\alpha \gamma} \rho_{\gamma}^{\; k} \left\{ p_{k}, \xi_{\beta } \rho^{\beta \, l} \right\}  p_{l} = 0 ,
\]
which is also equivalent to:
\begin{equation}
\rho^{i}_{\; \, \gamma} \left(G^{-1}\right)^{\gamma \delta} \rho_{\delta}^{\; \, j} = 0, \quad  \rho^{\alpha k} \left(\mathit{\Gamma}^{\beta}_{\; \; k \delta} \rho^{\delta l} + \partial_{k} \rho^{\beta l} \right) =0
\label{G_cond}
\end{equation}
This set of constraints on $G$, $\mathit{\Gamma}$ and $\rho$ is underdetermined. An ansatz must be made in order to solve it, depending on the deformation that one wants to discuss. In this paper we shall choose a (generically non-constant) anchor map that is null on forms, $\rho^{b i} = 0$, i.e. a projection onto tangent space up to rescaling by a function of the dilaton. In the discussion we will hint to other choices that can be investigated too. The option for $\rho$ that we pick up here, also implies that $G$ has a null lower right $d \times d$ block.

\paragraph{Canonical transformations:} Let us now discuss briefly a class of canonical transformations for this algebra, the inner automorphisms. These are degree-preserving morphisms and since the brackets have degree $-2$, the generators $h$ must  have degree $2$. This can be achieved by choosing for $h$ a function with a term linear in momenta and one quadratic in the degree $1$ coordinates:
\begin{equation}
h = v(x)^{i} p_{i} + \frac{1}{2} M^{\alpha \beta}(x) \xi_{\alpha} \xi_{\beta} \, .
\label{can_trafo}
\end{equation} 
When acting on an element of the algebra of functions of arbitrary degree $\mathcal{O}_{n}$ via Poisson bracket, the infinitesimal canonical transformations look like:
\[
\delta A = \{ h, A\}, \,  \quad A \in \mathcal{O}_{n}.
\]
The algebra of these generators $h$ closes as assured by degree counting and by the Poisson brackets themselves, since they produce a new function which is still linear in the momenta and quadratic in the $\xi$'s, see the relations \eqref{Pois_br_def}.
\[
\left[\delta_{1}, \delta_{2} \right] A = \{h_{1}, \{h_{2}, A \} \} - \{ h_{2}, \{ h_{1}, A \}\} = \{ \{ h_{1}, h_{2} \}, A \} = \{ h_{3} , A \} = \delta_{[h_{1}, h_{2}]} A\, . 
\]
The momenta generate diffeomorphisms and the term quadratic in the $\xi$'s generates local $\mathfrak{o}(d,d)$ transformations: $B \in \Lambda^{2}(T[1])$, $\beta \in \mathfrak{X}^{2}(T^{*}[1])$ and $A \in \text{End}(T^{*}[1] )$. One recognizes immediately that the latter are the symmetries that preserve the Courant algebroid pairing ($B$-transformations are actually automorphisms for Courant algebroids). The algebra of the canonical symmetries is hence the algebra of the gauge symmetry of the gerbe structure of $\left(TM \oplus T^{*}M , \rho, [\cdot, \cdot]_{\text{Dorf}}, \langle \cdot, \cdot \rangle\right)$.

These canonical transformations are of course derivations of the Poisson bracket. Considering the elements $K, W$ of $\mathcal{O}_{n}$ 
with $\{K, W\}=Z \in \mathcal{O}_{2n-2}$  
we have:
\[
\delta Z = \{h, Z \} = \{h, \{K,W\}\} = \{ \{h,K\}, W \} + \{K, \{h,W\} \} = \{ \delta K, W \} + \{ K, \delta W\} \, .
\]

\subsection{Derived structure}
\label{Cour_str}
In this part we want to focus on the outcome of the derived bracket construction, and describe the Courant algebroid corresponding to the deformation. 

As an example for the correspondence between the Poisson algebra of functions on $T^{*}[2]T[1]M$ with Hamiltonian $\Theta$ \eqref{TH_roy}, satisfying the master equation \eqref{master}, and the Courant algebroid $(TM \oplus T^{*}M, \rho, [\cdot, \cdot]_{\text{Dorf}}, \langle \cdot, \cdot \rangle)$ in section \ref{very_first} we have already discussed the canonical graded Poisson structure. Now we will investigate a modified Poisson structure together with the Hamiltonian \eqref{TH_roy}, whose master equation implies \eqref{G_cond}.\footnote{Later in this section we will ignore the flux terms in the Hamiltonian. Fluxes can be easily added later, they are essentially along for the ride and we suppress them here mostly for notational simplicity.}  The associated Courant algebroid with Dorfman bracket $[ \cdot, \cdot ]^{\prime}$, pairing~$\langle \cdot , \cdot \rangle$ and  anchor map $\rho$, as defined by the derived brackets \eqref{derived_brackets}, \eqref{der_br_func} and \eqref{pair} has the following structure:
\begin{enumerate}
\item The pairing is given by $G_{\alpha \beta}$: $\{\xi_{\alpha} , \xi_{\beta} \} = G_{\alpha \beta} = \langle \xi_{\alpha}, \xi_{\beta} \rangle$;
\item The anchor map follows from  $\{ \{ \Theta, f \} , \xi_{\alpha} \} = \rho(\xi_{\alpha}) f$ with $ f \in C^{\infty}(M)$;
\item The fully generalized Dorfman bracket $[ \cdot, \cdot ]^{\prime}$, written out in components, is given by
\begin{align}
[\xi_{\alpha}, \xi_{\beta} ]^{\prime} = \{\{ \Theta, \xi_{\alpha} \}, \xi_{\beta} \} &= \nabla_{\xi_{\alpha}} \xi_{\beta} - \nabla_{\xi_{\beta}} \xi_{\alpha} + G\left( \nabla \xi_{\alpha}, \xi_{\beta} \right) + \xi_{\gamma} C^{\gamma \nu \sigma } G_{\nu \beta} G_{\sigma \alpha}  \label{DORFF}\\
\; & =  \xi_{\gamma}\left(\rho_{\alpha}^{\; k} \mathit{\Gamma}^{\gamma}_{\; \; k \beta} - \rho_{\beta}^{\; k} \mathit{\Gamma}^{\gamma}_{\;\;  k \alpha} +G^{\gamma \epsilon} \rho_{\epsilon}^{\; k} \mathit{\Gamma}^{\delta}_{\; \; k \alpha} G_{\delta \beta}
+C^{\gamma \nu \sigma } G_{\nu \beta} G_{\sigma \alpha} \right) , \label{INDICES_D}
\end{align}
where the connection $\nabla : \Gamma(E) \times \Gamma(E) \rightarrow \Gamma(E)$ on the Courant algebroid  is induced by the connection of the graded symplectic manifold appearing in (\ref{Pois_br_def}) (as the usage of the same symbol aims to stress), 
\[
\nabla_{W} := \nabla_{\rho(W)}, \quad W \in \Gamma(E) ,
\] 
via the anchor map $\rho$ from generalized vector fields (which are equivalent to degree $1$ functions) to ordinary tangent vector fields (which are equivalent to 
 degree $2$ functions), $\rho: \Gamma(E) \rightarrow \Gamma(TM)$ (see also point 2 above). The connection is a bona fide extension to generalized vector fields and it satisfies the axioms for a generalized connection:
\[
\nabla_{W} fV = \left(\rho(W) f \right) V + f\nabla_{W} V, \quad \nabla_{fW} V = f \nabla_{W} V, \quad f \in C^{\infty}(M) , \; W,V \in \Gamma(E). \label{genconnection}
\]
\end{enumerate}

Before inspecting the deformation for the Courant algebroid to check once more that it is well-posed, let us comment on the presence of local generalized vielbeins $\text{E}$ relating the standard $O(d,d)$-invariant pairing to the new, completely general, metric $G$. Vielbeins are sections of a frame bundle for $TM \oplus T^{*}M$ and will be chosen so that the general metric $G$ could have some desired features. 
$\text{E}$ gives also a local isomorphism between exact Courant algebroids, therefore mapping a basis for sections $\Gamma(TM \oplus T^{*}M)$ of $(TM \oplus T^{*}M , \rho, [\cdot, \cdot]_{\text{Dorf}}, \langle \cdot, \cdot \rangle)$ to a basis for sections of $(E , \rho, [\cdot,\cdot]^{\prime}, \langle \cdot, \cdot \rangle^{\prime})$, and it is a homomorphism of the bracket too.

By na\"{i}vely modifying the graded symplectic structure one could potentially end up describing a wanna-be Courant algebroid that actually does not fulfill the defining conditions. As already explained, this will not happen as long as the Hamiltonian for the new deformed setting still has null Poisson bracket with itself, but it is nevertheless instructive to check directly that the quadruple $(E, \rho, [ \cdot, \cdot ]^{\prime}, \langle \cdot , \cdot \rangle^{\prime} )$ is really a Courant algebroid according to definition \eqref{Courant}. 
We will ignore the fluxes $C$ in $\Theta$ for now and consider only
\begin{equation}
\Theta \vert_{C=0} = \xi_{\alpha} \rho^{\alpha i} (x) p_{i},
\label{TH_noC}
\end{equation}
since one can observe that due to associativity the tensor $C$ is only responsible for the direct twisting of the bracket, by means of tensors with various symmetries, see \eqref{DORFF}. The choice \eqref{TH_noC} already reproduces the Courant algebroid bracket $[ \cdot, \cdot ]$, the anchor and the pairing. 
Then our previous computations of the master equation i.e. the formulas \eqref{Jacobi}, \eqref{2_ax} and \eqref{3_ax} are still true for $\Theta_{\vert C=0}$, again because of associativity. In particular, \eqref{Jacobi} imposes a rather obvious symmetry condition on the curvature $\text{R}$ of the induced connection, the algebraic Bianchi identity (which is trivially true for a vanishing Weitzenb\"{o}ck-type connection curvature)
\[
\left(\nabla_{[\beta} \nabla_{\gamma]} - \nabla_{\nabla_{[\beta} \xi_{\gamma]}} \right) \xi_{\alpha} + \text{perm}(\beta , \gamma, \alpha ) = \left(\text{R}^{\delta}_{\; \alpha \beta \gamma} + \text{R}^{\delta}_{\; \beta \gamma \alpha} + \text{R}^{\delta}_{\; \gamma \alpha \beta} \right) \xi_{\delta} = 0 ,
\]
where the shorthand notation $\nabla_{\beta} = \nabla_{\rho_{\beta}} = \nabla_{\rho(\xi_{\beta})}$ is used here, and another condition involving $G \equiv \langle \cdot, \cdot \rangle$ and the connection coefficients:
\begin{align*}
G\left( \nabla \xi_{[\beta} , \nabla_{[\gamma]} \xi_{\alpha]} \right) + G\left( \nabla \xi_{[\beta} , G\left( \nabla \xi_{\gamma]}, \xi_{\alpha} \right) \right) - \nabla_{G \left( \nabla \xi_{[\gamma \vert} , \xi_{\alpha} \right) } \xi_{\beta]} + \nabla_{[\beta} G\left( \nabla \xi_{\gamma]}, \xi_{\alpha} \right) \\
= G \left( \nabla \nabla_{[\beta} \xi_{\gamma]}, \xi_{\alpha} \right) + \nabla G\left( \nabla_{\alpha} \xi_{\beta}, \xi_{\gamma} \right)- \nabla_{\alpha} G \left( \nabla \xi_{\beta},\xi_{\gamma} \right) + \nabla_{G(\nabla \xi_{\beta}, \xi_{\gamma})} \xi_{\alpha} \, ,
\end{align*}
which is a disguised version of the previous symmetry condition on $\text{R} $.
Both remaining axioms do not require much more in order to be verified, since both \eqref{2_ax} and \eqref{3_ax} are simply equivalent to the statement that $\nabla$ is metric compatible, 
\begin{equation} \nabla G = 0, \label{mtrct} \end{equation}
which was already required in order to produce a Poisson structure with the brackets \eqref{Pois_br_def}.

To summarize, in this section we presented the general form of a graded Poisson structure \eqref{Pois_br_def}, which can be interpreted as a deformation of the canonical structure \eqref{Pois_br_def_canonical},  and the corresponding Hamiltonian, which was used to characterize the structure through the master equation. We then discussed infinitesimal degree-preserving canonical transformations of the graded Poisson structure. Finally, we derived the Courant algebroid on $TM \oplus T^{*}M$ arising from the derived brackets for the dg-symplectic manifold $T^{*}[2]T[1]M$ for such a general Poisson structure. The expression for the fully generalized Dorfman bracket of this algebroid, in equation \eqref{DORFF}, is found to involve a Weitzenb\"{o}ck-type connection for $ T^{*}[2]T[1]M$. This raises some interesting points that will be explored in the next section.

\section{Courant algebroid connection}
\label{middle}

In this part we would like to formulate some essential elements of differential geometry for the generalized tangent bundle $E$. For our particular way of constructing (super) gravity actions, we will not actually need all the results that are presented here in full generality, but they should be interesting in their own right and put the construction in a proper perspective.

 We will later apply the general formalism to the deformation given in~\eqref{Pois_br_def} with vanishing Weitzenböck-type curvature as before and the background fluxes $C$ ignored. For doing so and inspired by the structure of equation \eqref{DORFF}, we will start by looking for a suitable definition of a torsion tensor for $E$, in a coordinate-independent fashion. 

\subsection{Generalized torsion, Lie bracket and connection}
\label{sec:gentorsion}

A map $\text{T} : \Gamma(E) \times \Gamma(E) \rightarrow \Gamma(E)$ gives rise to a $2$-tensor if it is bilinear and antisymmetric: 
\[
\text{T}(U,fV) = f \, \text{T}(U,V), \quad \text{T}(U,V) = -\text{T}(V,U), \quad f \in C^{\infty}(M), \; U,V \in \Gamma(E).
\]
Moreover, it is a candidate for a torsion tensor, if it measures the failure of the antisymmetric pair of covariant derivatives on two generalized vector fields $\nabla_{U} V - \nabla_{V} U$  to close up to their commutator, that we could think of as a generalized Lie bracket $\llbracket \cdot , \cdot \rrbracket$. This  generalized Lie bracket shall have some properties determined by the behaviour of the connection $\nabla$ and the tensor $\text{T}$ themselves under $C^{\infty}(M)$-multiplication and by the antisymmetry of $\text{T}$. In previous works there have been attempts to employ the antisymmetric Courant bracket for these purposes, but that choice led to the addition of further correcting terms, mostly because it does not obey a Leibniz rule. We define instead:
\begin{defn} An  antisymmetric $\mathbb{R}$-bilinear bracket $\llbracket \cdot , \cdot \rrbracket : \Gamma E \otimes \Gamma E \rightarrow \Gamma E$, such that
\begin{equation}
\llbracket U, V \rrbracket = - \llbracket V, U \rrbracket, \quad \llbracket U, f V \rrbracket = \left(\rho(U) f \right) V + f \llbracket U, V \rrbracket, \;  f\in C^{\infty}(M),
\label{liEEE}
\end{equation}
holds for all $E$-sections $U,V$, is said to be a \textit{generalized Lie bracket}. (Note, that we do not \emph{require} a Jacobi identity, even though it may still hold in some generalized fashion.) 
\label{LIE}
\end{defn}
The definition does not uniquely determine a generalized Lie bracket, but any two choices are related by an $E$-section valued antisymmetric $2$-tensor. Existence of such a generalized Lie bracket is ensured for any Courant algebroid that has  a compatible generalized connection; see proposition~4.4 below. The ordinary Lie bracket of vector fields on $M$ gives a local condition $[\chi_i, \chi_j]_\text{Lie} = 0$ for a holonomic (coordinate) basis $\{\chi_i\}$. Vice versa, the choice of such a preferred basis can be used to define the bracket locally. We have a natural choice for such a preferred local basis, namely the one corresponding to the Darboux chart described before equation (\ref{symplectic}). We define this basis~$\{\xi_\alpha\}$ to be also holonomic in the generalized sense and consequently set $\llbracket \xi_{\alpha}, \xi_{\beta}\rrbracket =  0$.  With the help of the defining property \eqref{liEEE} we get the following explicit expression for the generalized Lie bracket of any two $E$-sections $U,V$ in a holonomic basis (local Darboux chart):
\begin{align}
\llbracket U, V \rrbracket &:=  \left( U^{\alpha}(x) \rho_{\alpha}^{\; i}(x) \partial_{i} V^{\beta}(x) - V^{\alpha}(x) \rho_{\alpha}^{\; i}(x) \partial_{i} U^{\beta}(x) \right) \xi_{\beta} \label{L_br} \\
& \equiv \rho(U) V - \rho(V) U . \notag
\end{align}
This expression is of course coordinate-choice dependent -- just as it is the case for the corresponding expression of the ordinary Lie bracket as a commutator of vector fields  -- but using again~\eqref{liEEE} it is straightforward to obtain expressions for other (non-holonomic) bases.
The bracket respects the defining conditions \ref{LIE} by definition. Its local expression  can be extended globally to all patches that cover the base manifold using~\eqref{liEEE} and the transition functions between patches of the underlying mani\-fold. The extension can easily be shown to be consistent on triple overlaps as long as the anchor map is globally well-defined.\footnote{A coordinate-choice independent expression for the generalized Lie bracket in terms of the canonical (generalized) solder form $\sigma$ that maps $TM\oplus T^*M \cong E$ to itself is $U.\sigma(V) -V.\sigma(U) -  d \sigma(U,V)$, where $U,V \in \Gamma (E)$ act on functions via the anchor map. Evaluated in a holonomic basis, the last term in the expression vanishes and \eqref{L_br} is recovered.} 
The above expression for the bracket resembles the commutator of generalized vector fields, in some sense analogously to the difference between the infinitesimal flow generated by the first generalized vector field on the second and vice versa.

For what concerns us, in this article we will ultimately only really use the generalized Lie bracket on ordinary tangent vector fields $X, Y \in \Gamma(TM)$, where it will correspond to the usual Lie bracket up to a rescaling by the dilaton that will appear in the anchor. 

\begin{exmp}[]{The original Dorfman bracket \eqref{DDD}, the standard Courant algebroid bracket, is partly given by $\llbracket \cdot, \cdot \rrbracket$ \eqref{L_br}. In fact, as one can easily check using $\rho \equiv \text{pr}_{TM}$,
\begin{align}
[e_{1}, e_{2}]_{\text{Dorf}} &=   \llbracket e_{1}, e_{2} \rrbracket + \langle \tilde{\text{d}} e_1, e_2 \rangle  \quad  e_{1}, e_{2} \in \Gamma(E)
\label{Dorfff}
\end{align}
where $\langle,\rangle \equiv \eta$ and we employed the derivative
$\tilde{\text{d}}: \Gamma(E)  \rightarrow \Gamma(E) \otimes \Omega^{1}(E) \cong \Gamma(E,E)$, which vanishes on the preferred basis $\{\xi_\alpha\}$, i.e. $\tilde{\text{d}} \xi_\alpha = 0$, and evaluated on a general section $e = e^\alpha(x) \xi_\alpha$ gives
\begin{equation}
 \tilde{\text{d}} e = \rho^*( \text{d} e^\alpha(x)) \xi_\alpha
\end{equation}
with $\text{d}$ the de Rham differential and $\rho^*$ providing the canonical embedding of the 1-form in $\Omega^{1}(E)$. The preferred basis is the one where the isomorphism of the exact Courant algebroid $E$ and $TM\oplus T^*M$ is trivial, i.e. $(\xi^\alpha) = (\chi_i,\theta^i) = (\partial_i, dx^i)$,  the fiber-wise metric is given by the natural pairing of $TM$ and $T^*M$  and the anchor map  is $\rho \equiv \text{pr}_{TM}$. It is the same preferred basis used in the definition of the generalized Lie bracket above.}
\label{exmpl}
\end{exmp} 

Finally, with the generalized Lie bracket \eqref{liEEE}, we are able to define  with
\begin{equation}
\text{T}(U, V) = \nabla_{U} V - \nabla_{V} U - \llbracket U,V \rrbracket 
\label{TORSSS}
\end{equation}
a genuine torsion tensor $\text{T} : \Gamma(E) \otimes \Gamma(E) \rightarrow \Gamma(E) $, with all the properties required (antisymmetry and linearity under multiplication of functions).

For a generalized connection $\nabla$, the following properties are required:
\begin{equation} 
\nabla_{W} fV = \left(\rho(W) f \right) V + f\nabla_{W} V, \quad \nabla_{fW} V = f \nabla_{W} V, \quad f \in C^{\infty}(M) , \; W,V \in \Gamma(E). \label{genconnection}
\end{equation}
In the remainder of this article we will often encounter expressions involving the connection contracted with a (generalized) vector field, i.e. ``with all indices lowered": $\langle \nabla_{W} U, V\rangle$. Furthermore, in section~\ref{body}, we will consider deformations involving a metric $g$, but not its inverse (the inverse will appear later), hence in that example we cannot expect to obtain Levi-Civita symbols directly, since they do involve the inverse metric. This suggests therefore to consider connection symbols of the first kind $\mathit{\Gamma}(W;U,V)$.
We shall stick to the convention for Christoffel symbols of the first kind in general relativity, where the upper index is lowered into the last slot: $\mathit{\Gamma}_{\gamma \alpha \beta} := \mathit{\Gamma}^{\lambda}{}_{\gamma \alpha}g_{\lambda\beta} $ and define connection coefficient on the generalized bundle $E$ with the appropriate tensorial and linearity properties:
\begin{defn}
A generalized connection symbol $\mathit{\Gamma}$ for a connection on the generalized tangent bundle $E $ is said to be \textit{of the first kind} if for 
a symmetric bilinear form $\langle \cdot, \cdot \rangle$, and $\rho : E \rightarrow TM$
\begin{equation}
\mathit{\Gamma}(W; f U,V) = \left(\rho(W)f \right) \langle U, V \rangle + f\mathit{\Gamma}(W;U,V)  , \quad \mathit{\Gamma}(W; U,fV) = f \mathit{\Gamma}(W; U,V) = \mathit{\Gamma}(fW;U,V),
\label{Propr}
\end{equation}
where $U,V,W$ are $E$-sections and $f$ is a smooth function.
\label{first_k}
\end{defn}
With the help of \emph{any} suitable non-degenerate symmetric bilinear form $\langle \cdot, \cdot \rangle$, a connection symbol of the first kind defines a generalized connection $\nabla$
(\ref{genconnection}) via
\begin{equation}
 \langle \nabla_{W} U, V\rangle := \mathit{\Gamma}(W;U,V) .
\label{EXAMPLE}
\end{equation} 
The proof is straightforward, but the important point is that there may be several choices of bilinear forms leading to  different  connections (some defined on subspaces). The relevant case will be $TM$, where an ordinary connection on vector fields is hence obtained when the arguments of $\mathit{\Gamma}$ are taken in the image of a generic (non-isotropic) splitting of the sequence \eqref{seq} $s : \Gamma(TM) \rightarrow \Gamma(E)$, $\rho \circ s = id_{TM}$, in the following way:
\begin{equation}
\mathit{\Gamma}\left(s(Z); s(X), s(Y) \right) =: \left \langle \nabla_{Z} X, Y \right\rangle_{\vert TM}, \quad X,Y,Z \in \Gamma(TM) \,,
\label{ord-con}
\end{equation}
then the operator $\nabla : \Gamma(TM) \times \Gamma(TM) \rightarrow \Gamma(TM)$ 
 fulfills all the requirements to be a connection because the symbols of the first kind $\mathit{\Gamma}$ do and composition with the splitting preserves those relations \eqref{Propr}. In order for $\nabla$ to be well-defined, the metric on the tangent space must be non-degenerate. Using a suitable splitting $s$, the tangent space metric can be obtained from the pairing on $E$ via  $\left \langle X, Y \right\rangle_{\vert TM} : = \langle s(X), s(Y) \rangle$, but then splitting $s$ must of course not be isotropic (because else $\langle s(X), s(Y) \rangle = 0, \; \forall \, X, Y \in \Gamma(TM)$). 

We shall denote the connection on the full generalized tangent bundle and the one on tangent space by the same symbol $\nabla$, because they both arise from the same connection symbols of first kind. Which one is in use should be immediately clear from the context.

\begin{propos} \label{propos}
Given a pairing  $\langle \,,\, \rangle$ (fiberwise metric on $E$) and an anchor map $\rho: E \rightarrow TM$, there is an interesting relation between the three structures defined before:
\begin{enumerate}
\item A bracket $[\,,\,]$ satisfying axioms 2 and 3 in definition \ref{Courant}, i.e. the axioms of a Courant algebroid except for the Jacobi identity, which may or may not hold,
\item a generalized Lie bracket $\llbracket \,,\, \rrbracket$ as in definition \ref{liEEE}, and
\item a connection $\mathit{\Gamma}(\;;\,,\,)$ as in  definition  \ref{first_k}, which is metric with respect to the pairing $\langle \,,\, \rangle$.
\end{enumerate}
Via 
\begin{equation}
\langle U, [V,W] \rangle = \langle U, \llbracket V,W \rrbracket \rangle + \mathit{\Gamma}(U;V,W) \label{triality}
\end{equation}
any two of the structures will define the third with all its desired properties. 

\vspace{1ex}

We have formulated this in the more general setting of a connection of type~1. Given an ordinary connection $\nabla$ and pairing, 
$\mathit{\Gamma}(U;V,W)$ can be replaced by $\langle \nabla_U V, W\rangle$.
\end{propos}

\noindent {\bf Proof.} We shall just sketch the main points: (i) Tensoriality in the first slot is obvious as can be seen by replacing $U$ by $f U$ with $f \in C^\infty(M)$ and considering the properties of the pairing and the connection. (ii) The failures of tensoriality of the two brackets in their last slots exactly mach: Both $[V,fW]$ and $\llbracket V,fW \rrbracket$ contribute an extra term $(\rho(V) f) \langle U, W\rangle$. These cancel and  imply that the connection is tensorial in the last slot and vice versa. For the first bracket this is equivalent to $(\ref{CLeibniz})$, which is directly related to axiom~3 in definition \ref{Courant}. (iii) The antisymmetry of the generalized Lie bracket $\llbracket V,W \rrbracket$ is related to axiom~2 in definition \ref{Courant} and metricity of the connection with respect to the pairing. To see this we set $W=V$. The properties to compare are the vanishing of $\llbracket V,V \rrbracket$, $\langle U, [V,V]\rangle = \frac 12 \rho(U) \langle V,V\rangle$ (axiom~2) and $\mathit{\Gamma}(U;V,V) = \frac 12 \rho(U) \langle V,V\rangle$, i.e.\ metricity of the connection with respect to the pairing. 
The general result follows by polarization. (iv) Finally, the non-tensorial terms in the middle slot are determined by replacing $V$ by $f V$. Axiom~2 and $(\ref{CLeibniz})$ imply extra terms $-(\rho(W) f) \langle U, V\rangle + (\rho(U) f)\langle V, W\rangle$ from the left hand side of (\ref{triality}). The first of these is matched by the one following from the definition of the generalized Lie bracket, the second one implies the connection 
property  $\mathit{\Gamma}(U;f V,W) = (\rho(U) f)\langle V, W\rangle + f \mathit{\Gamma}(U;V,W)$ and vice versa.  \hfill $\Box$ 

\vspace{1ex}

In conjunction with  a generalized connection -- that does not necessarily have to coincide with the one in~(\ref{triality}) -- the generalized Lie bracket is used to define torsion and curvature.

\subsection{Connection for the derived deformed Courant algebroid}
\label{sec:connection}

Consider now \eqref{DORFF} without fluxes $C$ and let $U^{\alpha} \xi_{\alpha} = U$ and $V^{\beta} \xi_{\beta} =V$:
\[
\left[ U, V \right]' = \nabla_{U} V - \nabla_{V} U + \langle \nabla U, V \rangle.
\]
Here $\nabla_U V = \nabla_{\rho(U)} V$ is the induced Weitzenb\"{o}ck-type connection introduced in section~\ref{Cour_str}.
The antisymmetric combination of covariant derivatives on the RHS becomes the torsion tensor \eqref{TORSSS} up to the generalized Lie bracket:
\begin{equation}
[ U,V ]^{\prime} - \llbracket U,V \rrbracket = \left\langle \nabla U, V \right\rangle + \text{T}( U,V ) .
\label{conn} 
\end{equation}
Contracting with $W$, we obtain a connection $\mathit{\tilde{\Gamma}}$  (of the first type), different than the induced Weitzenb\"{o}ck one for non-zero values of $\text{T}$
\begin{equation}
\left\langle W,  [U,V]^{\prime} - \llbracket U,V \rrbracket  \right\rangle = \left\langle V, \nabla_{W} U \right\rangle + \left\langle W, \text{T}( U,V )  \right\rangle  =: \mathit{\tilde{\Gamma}}(W;U,V) . 
\label{new_CoN}
\end{equation}
In fact, the LHS is clearly $\mathbb{R}$-linear in $U$ and tensorial elsewhere, hence $\mathit{\tilde{\Gamma}}$ are honest connection symbols of the first kind.
Remarkably it involves genuine generalized vector fields (i.e. both vector fields \textit{and} forms) everywhere, also in the direction along which the derivation is taken, because of the torsion tensor.

Furthermore, in light of proposition \ref{propos} here one can pick up any generalized Lie bracket, in particular the minimal one of example \ref{exmpl}. Then the connection $\mathit{\tilde{\Gamma}}$ is not anymore the Weitzenb\"{o}ck one but has non-trivial curvature. In components it reads
\[
\mathit{\tilde{\Gamma}}_{\alpha \beta \gamma} = \mathit{\Gamma}_{b \gamma \alpha} - \mathit{\Gamma}_{c\beta \alpha} + \mathit{\Gamma}_{a \beta \gamma},
\]
where the first index of $\mathit{\Gamma}$ is always anchored with $\rho$ and hence it can be labeled with the corresponding $d$-dimensional index.
$\mathit{\tilde{\Gamma}}$ will in general have non-zero curvature.

 Let us briefly comment at this point on the requirement of antisymmetry for torsion, since this is sometimes omitted by other authors but it plays an important role here: Metricity of $\nabla$ follows naturally from the Jacobi identity for a certain combination of derived brackets that imply the Courant algebroid axioms \eqref{2_ax} and \eqref{3_ax}. It also follows from the Jacobi identity for the underlying Poisson brackets. We have defined a new connection (of the first type) in \eqref{new_CoN}. Let us check its metricity directly:
 \begin{align*}
\rho(U) \langle V, W \rangle &=   \mathit{\tilde{\Gamma}}(U;V,W) +  \mathit{\tilde{\Gamma}}(U;W,V)\\
\, & = \langle W,  \nabla_{U} V \rangle + \langle \text{T}(V,W) , U \rangle + \langle V, \nabla_{U} W \rangle + \langle \text{T}(W,V), U\rangle, 
\end{align*}
 which is true, provided that torsion is antisymmetric and $\nabla$ is metric. \\

However, going back to the connection as defined from the connection symbols of first kind, since we are interested in a connection that returns (generalized) vectors rather than their duals, we could be tempted to perform an inversion with the full pairing $G^{-1} \equiv \langle \cdot, \cdot \rangle^{-1}$ on \eqref{EXAMPLE} to get
\[
\widetilde{\nabla}_{W} U = G^{-1} \mathit{\tilde{\Gamma}}(W ; U, \cdot).
\]
This yields a bona fide connection on the generalized tangent bundle, since $G$ is always non-degenerate if it is induced by a symplectic structure, as it is the case here. 
But for the particular deformation that we will consider, this approach will yield unphysical trivial results. We shall discuss this point in more detail in the context of curvature invariants for the manifold, subsection~\ref{Curvv}. 
In any case, physics should not be described on the doubled space -- it just plays an auxiliary role. We do need some kind of projection onto $TM$ and this can be achieved through a non-isotropic splitting $s$ as displayed in \eqref{ord-con}, to get a bona fide connection on the tangent bundle $\langle \widetilde{\nabla}_{Z}X, Y \rangle_{\vert TM}$.
The appropriate splitting turns out to be the natural embedding of $TM$ in $E$ up to a rescaling by the dilaton, which is needed to ensure $\rho \circ s = \text{id}$. 
Up to a similar rescaling, $\langle \cdot, \cdot \rangle_{\vert TM} =: G_{\vert TM}$, turns out to be the non-degenerate upper left $d \times d$ block of $G$ and 
\begin{equation} \widetilde{\nabla}_{Z} X = G_{\vert TM}^{-1} \mathit{\tilde{\Gamma}}(s(Z);s(X),s(\cdot) ) .\label{ON_vec} \end{equation}
Properties \eqref{Propr} remain true because composing with the splitting preserves naturally those relations. 

\subsection{Comparison with other definitions of torsion} 
Let us now go back to our definition of generalized torsion \eqref{TORSSS} 
\[
\text{T}(U, V) = \nabla_{U} V - \nabla_{V} U - \llbracket U,V \rrbracket  \,,
\] 
which is very similar to the standard definition of torsion,
and compare it with some other  proposals that have so far appeared in the literature. In the pioneering work by Coimbra, Strickland-Constable and Waldram, reference~\cite{Coimbra:2011nw}, torsion is the operator $\text{T}: \Gamma(E) \rightarrow \Gamma(\text{ad}F)$, for $\text{ad}F$ being the adjoint representation bundle associated to the bundle of frames. $\text{T}$ is given by the difference between a ``Dorfman derivative'' $\mathcal{L}$ taken with a covariant derivative, and a ``flat'' one (i.e. the Dorfman bracket):
\begin{equation}
\text{T}^{\nabla}(V) \cdot \alpha := \mathcal{L}^{\nabla}_{V} \alpha - \mathcal{L}_{V} \alpha,  \quad V \in \Gamma(E), 
\label{CSCW}
\end{equation} 
and $\alpha$ some tensor field in $E$. This formula replicates the definition of torsion in Riemannian geometry (where $\mathcal{L}$ is the Lie derivative). With the help of the pairing, $\text{T}^{\nabla}$ can be shown to belong to $\Gamma( \Lambda^{2} E \otimes E)$ and hence to be a tensor like its standard differential geometry counterpart.\\ 
In their work, $\nabla$ is compatible with a $O(d) \times O(d) \times \mathbbm{R}^{+}$ structure, where the positive reals take into account that the frame chosen is conformal. For the class of \emph{torsionless} covariant derivatives, the generic metric connection compatible with the generalized metric $\mathcal{H}$
\begin{equation}
\mathcal{H} = \begin{pmatrix}
g-Bg^{-1}B & -Bg^{1}\\
g^{-1}B & g^{-1}
\end{pmatrix}
\label{H_METRIC}
\end{equation}
and compatible with the conformal factor $\Phi$ as well ($\nabla \Phi = 0$) shall have $\text{T}^{\nabla}_{\alpha \beta \gamma}=0$. Then type~II closed strings bosonic effective action is recovered from the Ricci tensor, which is built with the commutator of covariant derivatives taking and acting on generalized vectors from both of the subspaces in which $\mathcal{H}$ restricts to a $O(d)$-invariant metric (or $O(1,d-1)$ metric in Lorentzian signature).

Before elaborating on the similarities of that approach with the present one, let us show a few more examples of torsion tensors. There are other natural ways to construct a $3$-tensor out of the data given by a Courant algebroid with compatible generalized connection, requiring in particular that, for the given connection $D$, $\text{T}^{D} \in C^{\infty}( \Lambda^{2} E \otimes E)$. This $3$-tensor is one of the candidates for a definition ``torsion" introduced by Gualtieri in \cite{Gualtieri:2007bq}. There, the Courant algebroid bracket employed in the definition is the skew-symmetric version. We shall rather stick to a torsion defined in the same fashion but with the Dorfman bracket instead (as in reference \cite{Jurco:2016emw}), and refer to it as Gualtieri torsion $\text{T}^{D}{}_{\text{\textbf{G}}}$ henceforth:
\begin{equation}
\text{T}^{D}{}_{\text{\textbf{G}}}(e_{3}, e_{1}, e_{2}) := \left\langle D_{e_{1}} e_{2} - D_{e_{2}} e_{1} - [e_{1}, e_{2}], e_{3} \right\rangle + \left\langle D_{e_{3}} e_{1}, e_{2} \right\rangle \, .
\label{Gualtieri}
\end{equation}
Here $D$ is compatible with the pairing. With this particular instance of the bracket \eqref{Gualtieri} is simply the contraction of \eqref{CSCW} with a third generalized vector, when $\alpha$ is a section of $E$.
Moreover, it can be shown to be also a special case of the definition of generalized torsion that we have given in this section: The Gualtieri torsion can be obtained by computing  the type-1 connection symbol $\mathit{\Gamma}$ (see definition~\ref{first_k} and eqn. \eqref{EXAMPLE}) from $D$ and using the result both in (\ref{triality}) to determine a generalized Lie bracket and in (\ref{TORSSS}) together with that Lie bracket to compute the torsion. The definitions of Gualtieri torsion and ours \eqref{TORSSS} are not identical, the latter being more general because one can also consider generalized Lie brackets and generalized connections not related by (\ref{triality})  to define bona fide generalized torsions via \eqref{TORSSS} that are not of Gualtieri type.

The connection that appears in the deformation of the graded Poisson structure 
 has in fact zero Gualtieri torsion with respect to the derived bracket based on a Hamiltonian without flux term. With flux term the Gualtieri torsion is equal to minus that flux. This is a direct consequence of the derived bracket construction, i.e. that construction is equivalent to imposing the torsion condition. In fact,
\begin{align*}
\text{T}^{\nabla}{}_{ \text{\textbf{G}}}(W,U,V) & = \left\langle \nabla_{U}V - \nabla_{V}U, W \right\rangle + \mathit{\Gamma}(W;U,V) - \langle \left[ U,V\right] , W \rangle \\
\; & \overset{!}{=} -C(W,U,V) \quad \text{using $[\cdot, \cdot], \langle \cdot , \cdot \rangle$, $\Theta_{C}$ and from } \eqref{new_CoN}.
\end{align*}

Another related definition, but for a covariant derivative not necessarily compatible with $\langle \cdot, \cdot \rangle$, was suggested by Alekseev and Xu in \cite{AntonPing:2001mt}:
\[
\text{T}(e_{1}, e_{2}, e_{3}) := \frac{1}{3} \, \text{cycl}_{123} \langle [e_{1}, e_{2} ]_{\text{Cou}}, e_{3} \rangle - \frac{1}{2} \, \text{cycl}_{123} \langle D_{e_{1}} e_{2} - D_{e_{2}} e_{1}, e_{3} \rangle \, .
\]
The authors discussed it in the context of Dirac generating operators for Courant algebroids. These operators, which are spin connections in the $\text{Spin}$ bundle, indeed arise naturally from the derived bracket construction. The quantization issue of the underlying $2$-graded symplectic manifolds was addressed in \cite{1410.3346}.


The Gualtieri torsion of the connection (\ref{new_CoN})
\begin{align}
\text{T}^{\widetilde{\nabla}}{}_{\text{\textbf{G}}}(W, U, V) &= \mathit{\widetilde{\Gamma}}(U;V,W) - \mathit{\widetilde{\Gamma}}(V;U,W) + \mathit{\widetilde{\Gamma}}(W;U,V) - \langle [U,V]', W \rangle \notag \\
\; & = \langle [V,W]', U \rangle - \langle [U,W]', V\rangle - \langle \llbracket V,W \rrbracket , U \rangle + \langle \llbracket U, W \rrbracket , V \rangle - \langle \llbracket U,V \rrbracket , W \rangle \,, \label{G_T}
\end{align}
as well as that of its tangent space ordinary counterpart \eqref{ON_vec} are in general non-zero unlike the one in the pioneering work \cite{Coimbra:2011nw}. The approach  of~\cite{Coimbra:2011nw} is hence manifestly different from the one presented here.

\subsection{Curvature invariants}
\label{Curvv}

In section~\ref{sec:connection} we have shown how to extract a generalized connection $\widetilde{\nabla}$ defined on the entire bundle from the derived connection of the first kind $\mathit{\tilde\Gamma}$. We have also provided a prescription on how to evaluate it on the tangent bundle only. Associated with these two options, there are also two curvature invariants for $\mathit{\tilde{\Gamma}}$. We will eventually only make use of the first one, but we shall briefly discuss both here.
\begin{enumerate}
\item Taking into account just the tangent bundle, \eqref{ON_vec} is the standard connection of Riemannian geometry. The Riemann curvature is thus the standard one $R \in \Gamma(TM \otimes \overset{\tiny{3}}{\bigotimes}T^{*}M )$:
\begin{equation}
\left[ \widetilde{\nabla}_{Z}, \widetilde{\nabla}_{Y} \right] X - \widetilde{\nabla}_{ \left([ Z,Y ] \right)} X= R( Z, Y) \, X.
\label{R_TM}
\end{equation}
The associated Ricci tensor is then the partial trace of $R$: $\text{Ric} \in \Gamma(T^{*}M \otimes T^{*}M )$, $\text{Ric}_{cb} = R^{a}_{\; \, cab}$.
This procedure may seem to not make full use of the rich structure of ``double geometry'', but it gives the correct physics and implements the curvature tensors for the tangent bundle as in usual Riemannian geometry. It also tremendously simplifies the calculations.
\item Considering the connection on generalized vector fields, extracted with the inverse of the full generalized metric $G$ instead, a generalized tensor in $\bold{R} \in \Gamma \left( E \otimes \overset{\tiny{3}}{\bigotimes} E^{*} \right)$ with the symmetries of the Riemann tensor can be defined in a similar way to \eqref{R_TM}:
\[
\left[ \widetilde{\nabla}_{W} , \widetilde{\nabla}_{V} \right] U - \widetilde{\nabla}_{\llbracket W,V \rrbracket} U = \bold{R}(W,V) \, U.
\] 
Since each of the indices can be vector or form valued, there are $2^{4} = 16$ Riemann curvatures, out of which only one corresponds to the standard differential geometry definition. The generalized Ricci tensor \[ \bold{R}\text{ic} \in \Gamma(E^{*} \otimes E^{*}) \] can be found again from the partial trace of this $\bold{R}$. There are $4$ of them and each one consists of the sum of $4$ Riemann curvatures with indices contracted.\\
In this way the curvature invariant is obtained for the full $E$-bundle. This would seem to lead to an alternative way to formulate an action, but in our case leads to trivial unphysical results. (Essentially, because the inverse of the metric $g$ on $M$ does not appear directly in the deformation that we consider in this article.)
\end{enumerate}

In the rest of the paper we will describe a particular choice for $G$ and $\nabla$. For the derived Riemann tensor and subsequently the Ricci tensor we will follow the prescriptions of the first point, namely they will be computed for the connection on the tangent bundle. $\text{Ric}$ will be contracted in a way that will reproduce the $10$-dimensional low-energy effective action for closed bosonic superstrings.

\section{Deformation with metric $g$, $2$-form $B$ and dilaton $\phi$}
\label{body}
\subsection{Deformed graded Poisson algebra}
 Here we will present a concrete example of a deformation. It will be based on $g+B$ (or equivalently $(g+B)^T = g-B$), i.e. the data that locally defines a generalized metric and also the string sigma model with background fields. Globally, $g$ and $H$ are relevant and we are dealing with a gerbe structure. By not introducing $g$ and $B$ separately, we can make sure that the relative prefactors of the terms in the action that we derive provides a non-trivial check of the construction. In this article we focus on an anchor map that is a projection onto tangent space up to a rescaling to accommodate a dilaton field. Other choices are possible and we briefly comment on them in the concluding discussion section \ref{DISC}.

By imposing that the anchor $\rho$ is a projector up to a rescaling, the metric $G$ shall have the lower block in the diagonal null, as discussed in the example after  equation~\eqref{G_cond}; subsequently the formulas there will then fix the expression for the connection symbols. Consider moreover that~$G$ deviates from the standard pairing by an overall conformal factor and a Riemannian metric~$g$ on the~$\chi$ coordinates. This Riemann metric $g : \Gamma(TM) \rightarrow \Gamma(T^{*}M)$ is extended to the full space of linear functions in the degree $1$ coordinates by composition with the anchor map $\rho$ and the embedding map $j: \Gamma(T^{*}M) \rightarrow \Gamma(TM \oplus T^{*}M)$ and through the isomorphism $\Gamma(TM) \cong \Gamma(T^{*}[2]M)$:
\[ j \circ g \circ \rho : \Gamma(T^{*}[1]M \oplus T[1]M) \rightarrow \Gamma(T^{*}[1]M \oplus T[1]M) .\]
Since the $p$-$p$ bracket is zero, 
the metric $G$ can be obtained from the standard $O(d,d)$ constant pairing $\langle \cdot, \cdot \rangle =: \eta$ by applying a vielbein $\text{E}$ based only on the Riemannian metric $g$, in matrix notation $G = \text{E}^{T} \eta \text{E}$. The vielbein is indeed a non-canonical (differentiable and invertible) change of the degree $1$ coordinates. In fact $\text{E}$ can also be seen as a choice of section in the associated bundle of frames to $T[1]M \oplus T^{*}[1]M$. 

The $B$-field can be introduced exploiting the structure of the underlying Courant algebroid. In fact, given that the pairing $\eta$ is invariant under $\mathfrak{so}(E) \simeq \Lambda^{2} \left(T[1]M \oplus T^{*}[1]M \right)^{*}$, it is always possible to add in the vielbein an antisymmetric $B \in \Lambda^{2} T[1]M$, which, as an application, is extendable to the full space $T^{*}[1]M \oplus T[1]M$ in the same way as $g$, i.e. $j \circ B \circ \rho$. The 2-form $B$ is defined up to $\Lambda$ transformations $B \mapsto B + d\Lambda$ (which is in fact part of the $O(d,d)$ symmetries). This freedom means that the vielbeins are defined locally (unless $H = dB$ is trivial) and are patched globally by $\Lambda$-transformations in addition to the usual transition functions. The overall structure is that of an abelian gerbe. Such a gerbe is one step up on the geometric ladder from a line bundle in the sense that now line bundles replace transition functions on the double overlaps of the cover of the underlying manifold.   See \cite{Aschieri:2002fq} for a discussion in a similar setting as the one we have here and for references to the original literature.

The local vielbein $\text{E}$ is then completely determined, and corresponds to:
\begin{equation}
\text{E} = \lambda \, \begin{pmatrix} \mathbbm{1} & 0 \\ g(x)-B(x) & \mathbbm{1}  \end{pmatrix} \, .
\label{vielbein}
\end{equation}
in which the conformal factor will be eventually set to some function $\exp f(\phi(x))$ for $\phi(x)$ a scalar field that will be interpreted as the dilaton of the gravity multiplet in supergravity.  With this vielbein we perform the local change of degree~1 coordinates that defines the deformation of the graded Poisson structure. This implies a local equivalence (isomorphism) of the corresponding deformed Courant algebroid to the standard exact Courant algebroid. This is however in general not a global equivalence, just like $B$ is not globally defined, but just a representative in a class locally defined by $H = dB$. 
We would furthermore like to point out that even a cohomologically trivial $H$ will in general be physically non-trivial just like a cohomologically trivial constant magnetic field will exert a force on charges.

The metric $G$ is hence
\begin{equation}
G = \text{E}^{T} \eta \text{E}= \lambda^{2} \begin{pmatrix} 2 g(x) & \mathbbm{1} \\ \mathbbm{1}  & 0 \end{pmatrix}  . 
\label{G}
\end{equation}
The local Weitzenb\"{o}ck-type connection \cite{Weitzenbock:1923efa} coefficients  $\mathit{\Gamma}^{\beta}_{\; \; i \alpha}$ are given by:
\[
\begin{aligned}
\xi_{\beta} \mathit{\Gamma}^{\beta}_{\; \; i \alpha} \, & = \left(\text{E}^{-1}\right)_{\gamma}^{\; \beta} \{p_{i}, \xi_{\beta} \text{E}^{\gamma}_{\; \alpha} \}\\
\, &= \xi_{\beta} \left(\text{E}^{-1} \right)_{\gamma}^{\; \beta}   \partial_{i} \text{E}^{\gamma}_{\; \alpha} \\
\, & = \xi_{\beta} \begin{pmatrix} \lambda^{-1} \partial_{i} \lambda \, \delta^{b}_{\; a}  & 0 \\ \partial_{i} \left( g - B \right)_{a b} & \lambda^{-1} \partial_{i} \lambda \, \delta_{b}^{\; a} \end{pmatrix} \, .
\end{aligned}
\]
where the Poisson bracket is the canonical one, and it is convenient to recall that $(\xi_{\alpha}) : = (\chi_{a}, \theta^{a})$. The connection is locally flat as it results from a local change of degree $1$ coordinates.
Moreover, as found in the end of subsection \ref{Cour_str}, the connection is clearly metric compatible with~$G$. (In order to avoid confusion, let us emphasize once more that these are local considerations and do not have any implications for the global properties of $M$ and its closely associated bundles beyond the requirement that a Riemannian metric $g$ and a closed 3-form $H$ must be globally well-defined.)

The expression for $\text{E}$, when written by making clear distinction between $\chi$ and $\theta$ coordinates, defines the following bracket in \eqref{Pois_br_def}:
\begin{equation}
\{ p_{i}, \chi_{a} \} = \lambda^{-1} \partial_{i} \lambda(x) \, \chi_{a} + \theta^{b} \partial_{i} \left(g(x)-B(x) \right)_{ba} \, , \quad \{p_{i}, \theta^{a} \} = \lambda^{-1} \partial_{i} \lambda(x) \, \theta^{a} \, .
\label{connec}
\end{equation}
The full (local) Poisson  structure that arises from the ansatz on $\rho$ and $G$ \eqref{G} is hence:
\begin{equation}
\begin{array}{ll}
\{p_{i}, x^{j} \} = \delta_{i}^{\; j}, & \{v, f \} = v(f) ,\\
\{\xi_{\alpha}, \xi_{\beta} \} = G_{\alpha \beta} , & \{U,V \} = G(U,V), \\
\{p_{i} , \xi_{\alpha} \} = \lambda^{-1} \partial_{i} \lambda \, \xi_{\alpha} + \theta^{b} \partial_{i} \left( g(x) - B(x) \right)_{ba} , & \{v, U\} = \nabla_{v}U , \\
\{p_{i}, p_{j} \} = 0 , & \{v, w \} = [v,w]_{\text{Lie}} \, ,
\label{br_GgB}
\end{array}
\end{equation}
with $(g-B)(U) = \left(g-B \right)(\cdot, U) = \left(g-B\right)_{a c}U^{c}$ and $ \nabla_{v} U = \lambda^{-1} v(\lambda) U + v \cdot j \circ (g-B)(\rho(U))$, where the dot denotes the action of the derivation $v$ on the element $j \circ (g-B)(\rho(U)) \in T^{*}[1]M \oplus T[1]M$. Elsewhere the notation is exactly as in \eqref{Pois_br_def}. As for the vielbein, transitions between the local Poisson structures on double overlaps of a cover of $M$ involve $\Lambda$-transformations. This reflects again the gerbe structure.

The closed non-degenerate symplectic form $\omega $ for this symplectic structure can be found from the canonical one by pull-back with $\text{E}$. In fact the change of degree $1$ coordinates $\text{E}$ is also a homomorphism of the Poisson algebra. Performing the substitution $\xi^{\text{\tiny{old}}} = \xi  \text{E}^{-1}$ in $d x^{i} \wedge dp_{i} + d \chi_{a} \wedge d \theta^{a} $ leads to: 
\begin{align}
\omega =& \, dx^{i} \left[ \lambda^{-3} \partial_{i} \lambda \, \partial_{j} B_{ab} \theta^{a} \theta^{b} \right] \wedge dx^{j} - dx^{i} \left[ \lambda^{-3} \partial_{i} \lambda \, \left(\chi_{b} -2 g_{ab} \theta^{a} \right)  + \lambda^{-2} \partial_{i} \left(g_{ab} - B_{ab}\right) \theta^{a} \right] \wedge d\theta^{b}  \notag \\
\, & \, - dx^{i} \left[ \lambda^{-3} \partial_{i} \lambda \, \theta^{a} \right] \wedge d\chi_{a} +d\theta^{a} \left[ \lambda^{-2} \right] \wedge d \chi_{a} - d \theta^{a} \left[ \lambda^{-2} g_{ab} \right] \wedge d\theta^{b} + dx^{i} \wedge dp_{i} \, . \label{symp}
\end{align}
The inversion of the Poisson bivector agrees with this result, as well as \eqref{sy_def} for the metric and connection employed here.

At this point we should unveil the Hamiltonian $\Theta_{\vert C=0} = \rho^*(dx^i) p_i = \xi_{\beta}\left(G^{-1} \right)^{\beta \alpha} \rho_{\alpha}{}^{ i} p_{i}$ compatible with the symplectic form \eqref{symp}. It is given by:
\begin{equation}
\Theta \vert_{C=0} = \lambda^{-1} \theta^{i} p_{i} . \label{HAAM}
\end{equation}
The structure equation for the Hamiltonian \eqref{master} still holds, because it now involves the deformed commutation relations \eqref{br_GgB} rather than the CCR. The direct computation results again in 
\[
\{ \lambda^{-1} \theta^{j} p_{j}, \lambda^{-1} \theta^{i}  p_{i} \} = 0 \,,
\]
because the derivatives on the conformal factor $\lambda^{-1}$ cancel with terms in the connection coefficients. 
Moreover, the composition of maps $\rho \circ G^{-1} : \Gamma(E^{*} ) \rightarrow \Gamma(TM)$ that appears in $\Theta$ can be seen to be: 
\begin{equation}
\left(G^{-1}\right)^{a \beta} \rho_{\beta}{}^{ i} = 0, \quad \left(G^{-1}\right)_{a}{}^{\beta}\rho_{\beta}{}^{ i}= \lambda^{-1} \delta_{a}{}^{i} .
\label{duall}
\end{equation}

\subsection{Derived structure}
Let us describe in detail how the derived structure for this particular deformation induced by the local vielbein $\text{E}$ \eqref{vielbein} looks like.
We will keep identifying sections of the algebroid with the algebra of functions of degree $1$ of $T^{*}[2]T[1]M$ by slight abuse of notation.  In this way the vielbein $\text{E}$ is also interpreted as a smooth map between sections of $(TM \oplus T^{*}M, \rho, [\cdot, \cdot]_{\text{Dorf}}, \langle \cdot, \cdot \rangle )$ and sections of $(E, \rho, [ \cdot, \cdot]^{\prime}, \langle \cdot, \cdot \rangle^{\prime} )$, which is the Courant algebroid corresponding to the deformed Poisson algebra \eqref{br_GgB}. 
The new pairing (now denoted with a prime) is directly:
\begin{equation}
\{e_{1}, e_{2} \} = \langle e_{1}, e_{2} \rangle^{\prime} = \lambda^{2} \langle e_{1}, e_{2} \rangle + 2 \lambda^{2} g \left( e_{1}, e_{2} \right) \;\; \; \forall \, e_{1,2}= e_{1,2}^{\; \; \; \, \; \alpha} \xi_{\alpha}   \in \Gamma(E)  . 
\label{Cou_pair}
\end{equation}
The new pairing has a reduced symmetry group with respect to  $\eta$, namely is invariant under $\left( O(d) \times O(d) \right) \ltimes e^{B}$. This is to be compared with the approach usually taken in Generalized Geometry, where the metric considered in order to define a generalized connection is $O(d) \times O(d) $ invariant. Furthermore the maximal isotropic subspaces for \eqref{Cou_pair} are no longer $TM$ and $T^{*}M$, as they were for the $\eta$ pairing, and they are neither the subbundles on which the metric $\mathcal{H}$ \eqref{H_METRIC} restricts to separate euclidean metrics of signature $d$ and $-d$ respectively. The maximal isotropic subspaces  $N_{0}$ and $N_{1}$ for \eqref{Cou_pair} are instead:
\begin{align}
N_{0} = \{ (X, \sigma) \in E \, \vert \, \sigma= - (g(X) - B(X))  \} \, , \quad N_{1} = T^{*}M \, .
\label{isotr}
\end{align}
Their dimension  is  $\dfrac{1}{2} \text{dim} (E)$. The fact that $E$ splits into $N_{0}$ and $N_{1}$ is one of the interesting different features of this model, in contrast to the standard literature on the topic. 

The definitive anchor map for the Courant algebroid is in components the very same $\rho_{\beta}{}^{i}$ highlighted in expression \eqref{duall} in the previous subsection. Let us eliminate the metric $G$ from there and reproduce the resulting $\rho : \Gamma(E) \rightarrow \Gamma(TM)$ here for $e=X+\xi$:
\begin{equation}
\rho(e) = \lambda X.
\label{Cou_anch}
\end{equation}
We can also immediately state that for an exact Courant algebroid the embedding $j$, regardless of the particular choice, is forced to have this scaling behaviour w.r.t. the embedding $j^{\text{\tiny{old}}}$ for the standard Courant algebroid:
\begin{equation}
j( \sigma) = \lambda^{-1} j^{\text{\tiny{old}}}(\sigma) \, , \quad \forall \, \sigma \in \Gamma(T^{*}M)  \, .
\label{Cou_emb}
\end{equation}
For the Poisson brackets \eqref{br_GgB} the Courant algebroid bracket in \eqref{DORFF} becomes:
\begin{align}
[e_{1}, e_{2} ]^{\prime} = &\lambda [e_{1}, e_{2} ]_{\text{Dorf}} + \lambda^{-1}\rho(e_{[1}) \lambda \, e_{2]} +  \rho(e_{[1}). j  \left(g-B \right)\left(\rho( e_{2]} )\right)\notag \\
\, & +  \langle \rho(\cdot). j\left(g-B\right) \left( \rho(e_{1}) \right) , e_{2} \rangle + \left(\langle e_{1}, e_{2} \rangle+ 2 \, g( e_{1}, e_{2} ) \right) \, \lambda^{-1} \iota_{\rho(\cdot)} d \lambda \, .
\label{COU_br}
\end{align}
The parenthesis on the indices means antisymmetrization without $\frac{1}{2}$ factor,  and the dot in $\rho(e_{n}) . \\ j \left(g-B \right) \left(\rho(e_{m}) \right)$ denotes the action of the derivation by the anchored $e_{n}$ on $j \circ \left(g-B\right) \circ \rho$, which hence neglects $e_{m}$. 

This modified Dorfman bracket can be rewritten in a way that resembles the canonical one. To do so, one should notice that $\text{E}$ being a homomorphism of the Poisson bracket, is also a homomorphism of the Dorfman bracket
\begin{align}
[e_{1}, e_{2} ]^{\prime} = \text{E}^{-1} \left( \left[  e_{1}^{\; a} \text{E}_{a}, e_{2}^{\; b} \text{E}_{b} \right] \right) = & \lambda [e_{1}, e_{2} ]_{\text{Dorf}} - \lambda (g-B) ([e_{1}, e_{2}]) + \lambda^{-1} \rho(e_{[1}) \lambda \, e_{2]}  \notag \\
\, &  + \mathcal{L}_{\rho(e_{1})} (g-B)(e_{2})   - \iota_{\rho(e_{2})} d(g-B)(e_{1})  \notag \\
\, & + \left(\langle e_{1}, e_{2} \rangle  + 2 \,  g (e_{1}, e_{2})  \right) \, \lambda^{-1} \iota_{\rho(\cdot)} d \lambda \, . 
\label{br_to_Kos}
\end{align}
Here again the derivation $\rho(e)$ does not act on the generalized vector field on which $g-B : T^{*}M \rightarrow TM$ is valued. \\
As stressed in subsection \ref{Cour_str}, the axioms of the Courant algebroid definition \ref{Courant} are proved for the pairing in \eqref{Cou_pair}, the anchor map in \eqref{Cou_anch} and the bracket in \eqref{COU_br}. Thus these objects yield a well-defined Courant algebroid. For a further check one could also prove this by using the observation that $\text{E}$ is a homomorphism between the two Courant algebroid brackets (see \eqref{br_to_Kos}), and the Leibniz rule and the Jacobi identity for $(TM \oplus T^{*}M, \rho, [ \cdot, \cdot ]_{\text{Dorf}}, \langle \cdot, \cdot \rangle)$.

To conclude, let us repeat again that the derived structure associated to the Poisson algebra in \eqref{Pois_br_def} with Hamiltonian \eqref{HAAM}, when the fluxes are set to zero, is that of a Courant algebroid. Since the local vielbein $\text{E}$ in \eqref{vielbein} maps the canonical Poisson structure to the deformed one, at the same time $\text{E}$ maps the Courant algebroid $(TM \oplus T^{*}M, \rho, [\cdot, \cdot]_{\text{Dorf}}, \langle \cdot, \cdot \rangle)$ to the deformed $(E, \rho, [\cdot, \cdot]^{\prime}, \langle \cdot , \cdot \rangle^{\prime})$.
While doing so, the connection introduced in the deformed Poisson bracket setting becomes an induced Courant algebroid connection. We will extensively discuss it after presenting, in the next paragraph, a further application of this modified structure.

\subsubsection{Courant $\sigma$-model} 
 As a straightforward application we can immediately compute the Courant $\sigma$-model corresponding to the deformation. This $\sigma$-model is the AKSZ action functional for a theory of extended objects with a Courant algebroid as underlying symplectic algebroid. 
\begin{equation}
\mathcal{S}_{\text{Cou}} = \int_{\Sigma} \Pi_{i} \wedge d x^{i} + \frac{1}{2} \eta_{\beta \gamma} \, \alpha^{\beta} \wedge d \alpha^{\gamma} - h(x)^{i}_{\; \; \beta} \alpha^{\beta} \wedge \Pi_{i}  + \frac{1}{6} C_{\beta \gamma \delta} \, \alpha^{\beta} \wedge \alpha^{\gamma} \wedge \alpha^{\delta} \, .
\label{sigma}
\end{equation}
The manifold $\Sigma$ is the $3$-dimensional worldvolume of a membrane, $h \in E^{*} \otimes TM$ for $M$ smooth dg-manifold and for $E := TM \oplus T^{*}M$, $C \in \overset{\tiny{3}}{\bigotimes} \, E^{*}$, $i = 1, \dots, d$ and $\beta= 1, \dots, 2d$. The functional \eqref{sigma} describes a symplectic structure on the space $\text{Maps}(\Sigma, M)$.  In this sense the integral is over the $0$-degree Hamiltonian for the symplectic structure; $x^{i}$ are the coordinates, $\Pi_{i}$ the momenta $2$-forms and $\alpha^{\beta}$ are the $1$-form associated to the degree $1$ coordinates.
The graded variables in the previous section are in fact the point particle analogue of these other ones, according to the following formal assignation:
\[
\left(x, \xi, p \right) \, \leftrightarrow \left( x, \alpha, \Pi \right) \, .
\]
Also their grading and hence their odd/even parity can be naturally interpreted in terms of ($0$-forms,) $1$-forms and $2$-forms.
The deformed Courant $\sigma$-model with $(TM \oplus T^{*}M, \rho, [ \cdot, \cdot ]', \langle \cdot , \cdot \rangle')$ as underlying symplectic algebroid is found by pullback, thus one needs to apply the vielbein map $\text{E}$
\[
\alpha \rightarrow \text{E} \alpha \, .
\]
By making clear distinction between shifted $1$-forms $\theta$ and shifted vector fields $\chi$ and using as $h$ (prior to the deformation) the projector, \eqref{sigma} becomes
\begin{align}
\mathcal{S}_{\text{Cou}}^{\prime} = \int_{\Sigma} & \Pi_{i} \wedge dx^{i} + \frac{1}{2}\lambda^{2} \partial_{i} B_{ab} \, \theta^{a} \wedge dx^{i} \wedge \theta^{b} +  \lambda^{2} g_{ab} \theta^{a} \wedge d\theta^{b}  \notag \\
\, & + \frac{1}{2} \lambda^{2} \left( \theta^{a} \wedge d \chi_{a} + \chi_{a} \wedge d \theta^{a} \right) - \theta^{i} \wedge \Pi_{i} + \frac{1}{6} \lambda^{3} H_{abc} \, \theta^{a} \wedge \theta^{b} \wedge \theta^{c} \notag \\
\, &  + \frac{1}{6} \lambda^{3} f_{ab}^{\; \; \;\; c} \theta^{a} \wedge \theta^{b} \wedge \hat{\chi}_{c} + \frac{1}{6} \lambda^{3} Q_{a}^{\; \; bc} \theta^{a} \wedge \hat{\chi}_{b} \wedge \hat{\chi}_{c}  + \frac{1}{6} \lambda^{3} R^{abc} \hat{\chi}_{a} \wedge \hat{\chi}_{b} \wedge \hat{\chi}_{c} \, ,
\end{align}
where we have used the shorthand notation $\hat{\chi}_{a} := \chi_{a} + (g-B)_{ba} \theta^{b} $.

Interestingly, the momentum $P$ canonically associated to the $x$ coordinate is now $P_{i} = \Pi_{i} - \lambda^{2} \partial_{i} B_{ab} \theta^{a} \theta^{b}$.  But this comes with no surprise if one adopts the viewpoint that the modification of the symplectic and Poisson structure can be considered a way to introduce interactions. 

\subsection{Connection and curvature}
\label{CONN_gBphi}
In this paragraph we readily adapt the general formulas for the connection \eqref{new_CoN}, \eqref{ON_vec} to the case under consideration. The result for the connection of the generalized tangent bundle is briefly discussed, as the main focus is on the connection for tangent space.

First of all a generalized connection provides a covariant derivative on generalized vector fields. The covariant derivative, by definition, behaves in the following way:
\begin{equation}
\widetilde{\nabla}_{W} U = \xi_{\gamma} \rho(W)^{a} \partial_{a} U^{\gamma} +  \mathit{\tilde{\Gamma}}^{\gamma}{}_{\alpha \beta} \xi_{\gamma} W^{\alpha} U^{\beta } ,  \quad  W= \xi_{\alpha} W^{\alpha}, \, U = \xi_{\beta} U^{\beta} \in \Gamma(E).
\label{cov_der}
\end{equation}
As expected, the differential operator comes with a factor of $\lambda$ in front of the partial derivative, carried by the anchor. The covariant derivative on vector fields, instead, will not have such factor as embedding vector fields with the splitting $s$, which is the left inverse of $\rho$, will take it away.\\
To see now what is the local expression of the connection derived from the vielbein $\text{E}$ we can display the deformed Courant algebroid Dorfman bracket in the local coordinate basis, i.e. in place of $e_{1}, e_{2}$ in \eqref{COU_br} we will plug in the coordinate basis $(\partial_{a}, dx^{a}) =: \xi_{\alpha}$. The same will be done for the Lie bracket, which is then identically zero as its definition \ref{LIE} is formulated in such frame.

The components of $\widetilde{\nabla}$ (i.e. the connection coefficients of the first kind, see definition \ref{first_k} and equation \eqref{new_CoN}) can be collected in the following way:
\begin{align}
\langle \widetilde{\nabla}_{\xi_{\gamma}} \xi_{\alpha}, \xi_{\beta} \rangle^{\prime} & = \langle \widetilde{\nabla}_{\partial_{c}} \xi_{\alpha} + \widetilde{\nabla}_{ d x^{c}} \, \xi_{\alpha}, \xi_{\beta} \rangle^{\prime} \notag \\
= & \lambda^{3}  \begin{pmatrix} 2 \Gamma^{\text{\tiny{L.C.}}}_{\; \; \; \; \; bca}  + H_{bca} + 2 \lambda^{-1} \left(\partial_{(c} \lambda(x) \, g_{a)b} - \partial_{b} \lambda(x) \, g_{ca} \right) & \; & \lambda^{-1} \partial_{(c} \lambda(x) \, \delta_{a)}{}^{ b} \\
\lambda^{-1} \left(\partial_{c} \lambda(x) \, \delta_{b}{}^{a} - \partial_{b} \lambda(x) \, \delta_{c}{}^{ a}  \right)  & \; & 0 \end{pmatrix}  \notag \\ 
\, & + \lambda^{3} \begin{pmatrix}
 \lambda^{-1} \partial_{[a} \lambda(x) \, \delta_{b]}{}^{c} & 0 \\ 
0 & 0
\end{pmatrix} .
\label{comps}
\end{align}
In the first block of the first matrix , $\widetilde{\nabla}_{ \partial_{c}}^{(1,1)}$, the particular combination of derivatives on $g+B$ has reproduced the Christoffel symbols of first kind $\Gamma^{\text{\tiny{L.C.}}}$ plus the Neveu-Schwarz tensor $H$, that here corresponds directly to the exterior derivative of $B$, $H = dB$. Everywhere else, i.e. $\widetilde{\nabla}_{\partial_{c}}^{(1,2)}$, $\widetilde{\nabla}_{ \partial_{c}}^{(2,1)}$ and $\widetilde{\nabla}_{d x^{c}}^{(1,1)}$, the connection depends only on derivatives on the conformal factor, which appears because of metricity with $\langle \cdot , \cdot \rangle^{\prime} = G$. 

For the components of the connection for tangent space we can substitute $\xi_{\gamma}, \xi_{\alpha}$ and $\xi_{\beta}$ with $s(\partial_{c}), s(\partial_{a})$ and $s(\partial_{b})$ respectively, where the non-isotropic splitting $s$, $\rho \circ s = \text{id}$, is chosen to be the embedding up to the inverse of $\lambda$:\footnote{Alternatively, one can use a projection onto tangent space without a rescaling by $\lambda^{-1}$. This will shift around some $\lambda$-factors in the formulas, but eventually it gives the same end-result as the arguably slightly more elegant approach presented here.}
\begin{equation}
s: \Gamma(TM) \mapsto \Gamma(E) , \quad s(X) = \lambda^{-1}X.
\label{s}
\end{equation}
The induced metric on tangent space, in light of \eqref{s}, is
\[
\langle s(X), s(Y) \rangle' = 2g(X,Y).
\] 
Employing formulas \eqref{ord-con} and  \eqref{ON_vec}, for the present deformation the following connection symbols on vector fields arise:
\begin{align}
2g\left( \widetilde{\nabla}_{\partial_{c}} \partial_{a}, \partial_{b} \right) & = \langle \widetilde{\nabla}_{s(Z)} s(X) , s(Y) \rangle' \notag \\
\; & = \left\langle  \lambda^{-1} \widetilde{\nabla}^{(1,1)}_{\lambda^{-1} \lambda^{-1} \partial_{c}} \partial_{a} , \partial_{b} \right\rangle' - \lambda^{-3} \partial_{c} \lambda \, \langle \partial_{a}, \partial_{b} \rangle', \notag 
\end{align}
We hence get the following values for the connection symbols:
\begin{equation}
\mathit{\tilde{\Gamma}}^{b}{}_{ca} =  \left( \Gamma^{\text{\tiny{L.C.}} }\right)^{b}_{\; \; ca} + \frac{1}{2} H^{b}_{\; \; ca} + \lambda^{-1}\left(\partial_{a} \lambda(x) \, \delta_{c}^{\; b} - \partial^{b} \lambda \, g_{ca} \right) .
\label{Nabla}
\end{equation}
This expression is surely interesting. A first thing that can be discussed is its Gualtieri torsion $\text{T}^{\widetilde{\nabla}}{}_{\text{\textbf{G}}}$. Using formula \eqref{G_T} and \eqref{COU_br} for the bracket, $\text{T}^{\widetilde{\nabla}}{}_{\text{\textbf{G}}}$ is non-zero and in fact 
equal to $H=dB.$ The torsion part of the connection \eqref{Nabla} and the contributions due to metricity will be analyzed also in a different fashion in subsection \ref{TORS_CONN}.
Let us also point out that the connection depends on $B$ only via $H$ and is hence invariant under $\Lambda$-transformations as anticipated.

In the following section we will highlight the natural appearance of a generalized Koszul formula.  

\subsubsection{Koszul formula}
\label{KOSZ}
Let us underline here a peculiar property of the connection which is immediate if we rather perform the computation differently. Focusing on pure vector fields $(U = X, V = Y, W=Z)$, in equation \eqref{new_CoN}, with $[ \cdot, \cdot ]^{\prime}$ given by \eqref{br_to_Kos} and $\llbracket \cdot, \cdot \rrbracket$ as before \eqref{L_br}, if we rather express the LHS before the first equality employing some basic properties of the Lie derivative $\mathcal{L}$ in the expression of the Dorfman bracket, namely the product rule and Cartan formulas, it will yield the Koszul formula for an antisymmetric metric $\mathcal G := g-B$ and a scalar field $\phi(x)$. This generalization of the Koszul formula is:
\begin{align}
g\left( \widetilde{\nabla}_{Z} X, Y \right) = & \frac 12 \Big( X.\mathcal G (Z,Y)
- Y.\mathcal G(Z,X) + Z.\mathcal G(Y,X) \nonumber\\
&\quad +\mathcal G([Y,Z],X) - \mathcal G([X,Z],Y) - \mathcal G(Z,[X,Y])\Big) \nonumber \\
& +\lambda^{-1}(X.\lambda) g(Y,Z) - \lambda^{-1}(Y.\lambda)g(X,Z)
\label{KSZL}
\end{align}  
From \eqref{KSZL}, it is easy read off the connection in components: 
\[
\mathit{\widetilde{\Gamma}}^{j}_{\; \, ki}  =  \frac{1}{2} \, g^{mj} \Big(\partial_{i} (g+B)_{mk} - \partial_{m}(g+B)_{ik} + \partial_{k} (g+B)_{im} \Big)   + \lambda^{-1} \left(\partial_{i} \lambda \, \delta^{j}_{\; \; k} - \partial^{j} \lambda \, g_{ik}  \right)  \, .
\]
Notice that we flipped indices picking up the opposite sign for $B$ everywhere. These connection coefficients are in agreement with \eqref{Nabla}, because clearly
\[
g^{mj} \left(\partial_{i} B_{mk} - \partial_{m} B_{ik} + \partial_{k} B_{im} \right) = g^{mj} H_{imk} \, ,
\]
while derivatives on the metric built up the Levi-Civita connection.

To our knowledge, this is a totally new generalization of the Koszul formula to a non-symmetric metric and a conformal factor. It solves ambiguities of other non-symmetric metric approaches to gravity. It enjoys also the gauge symmetry $B \mapsto B + d\Lambda$ reflecting the underlying abelian gerbe structure. Moreover, unlike in the standard GR case where the formula is derived from three combinations of the metricity condition obtained by permutations of the vector fields and assuming a torsion condition, here the Koszul formula is a direct result of the metric connection's definition \eqref{conn}, and in a broader sense of the derived brackets.

\subsubsection{Metric connection with torsion}
\label{TORS_CONN}
Let us also comment briefly on another viewpoint on the connection for tangent vectors.
The starting point of the discussion was a graded Poisson algebra endowed with a Weitzenb\"{o}ck-type connection, which is metric with respect to $G$ and pure torsion (where the torsion tensor takes values in the generalized tangent space). Then in the derived bracket construction the appearance of the torsion tensor \eqref{TORSSS} on a different combination of generalized vectors was noticed, and a new connection $\widetilde{\nabla}$ was defined using it, because the difference between two metric connections is necessarily a tensor.

In the standard Riemannian geometry of $TM$, a metric connection is obtained from the unique metric torsion-free connection with respect to the induced metric from $G$ to tangent space, i.e. $ 2 \lambda^{2} g$, (hence the Levi-Civita one) by adding the contorsion tensor $\text{K}$. 
For a torsion tensor on vector fields only, $\text{T} \in \Gamma(\Lambda^{2}T^{*}M \otimes TM)$, the contorsion is the tensor $\text{K} \in \Omega^{3}(M), \,  \text{K}_{kij} := \frac{1}{2} \left( \text{T}_{kij} - \text{T}_{ijk} + \text{T}_{jki} \right)$. Given that in the local coordinate basis the torsion $\text{T}_{kij}$ is 
\begin{align*}
\text{T}_{kij} &= \partial_{k} \left(g+B\right)_{ij} + \partial_{[i} \left(g+B\right)_{j]k} + \lambda^{-1} \partial_{[i} \lambda g_{j]k},
\end{align*}
$\text{K}$ becomes directly
\begin{equation}
\text{K}_{kij} = H_{kij} + \lambda^{-1} \partial_{[i} \lambda g_{j]k} \, .
\end{equation}
The derivatives on the metric drop out as expected.

Hence, renaming
\[
2 \lambda^{-1} \left( \partial_{i} \lambda(x) \, g_{jk} - \partial_{j} \lambda(x) \, g_{ik} \right) =: \alpha_{kij} \quad \alpha \in \Gamma(\Lambda^{2} T^{*}M \otimes T^{*}M) \, ,
\]
the following compact expression is obtained
\begin{equation}
\widetilde{\nabla} = \nabla^{\text{\tiny{L.C.}}} + \frac{1}{2} g^{-1}H + \frac{1}{2} g^{-1} \alpha \, .
\label{onTM}
\end{equation}

In the discussion after the definition and the derivation of a connection in the most general setup for the derived structure \eqref{new_CoN} we already highlighted similarities and differences with other works on the topic. In the specific case of the metric $G$ under consideration, we would also like to stress that the mixed symmetry form $\alpha$ as in \eqref{onTM} reminds us of the $1$-form valued endomorphism (\emph{Weyl term}) in the unique decomposition of the difference between two connections for $E$ with the same torsion, see \cite{Garcia-Fernandez:2016ofz} and \cite{Garcia-Fernandez:2015hja}. The endomorphism discussed here is the vielbein $\text{E}$ and $\alpha$ takes into account the variation of a metric connection with fixed torsion upon a conformal rescaling of the metric with $\lambda$ factors.

Now we can proceed with computing the Riemann tensor $R^{b}_{\; a c d}$ for the connection on the tangent bundle \eqref{Nabla}. Following therefore definition \eqref{R_TM}:
\begin{align}
R^{b}_{\; a cd} = & \left(R^{\text{\tiny{L.C.}} }\right)^{b}{}_{ acd}\, + \frac{1}{2} \nabla^{\text{\tiny{L.C.}}}_{[c} H^{b}{}_{ d]a} + \frac{1}{4} H^{b}{}_{ [ c \vert l} H^{l}{}_{ \vert d] a} +\nabla^{\text{\tiny{L.C.}}}_{[c \vert}  \lambda^{-1}  \left( \partial_{a} \lambda(x) \, \delta_{ \vert d]}{}^{b} - \partial^{b} \lambda(x) \, g_{ \vert d]a}\right) \notag \\
\, &  + \lambda^{-2} \left( \partial_{l } \lambda(x) \, \delta^{b}{}_{ [c \vert} - \partial^{b} \lambda(x) \, g_{l [c \vert} \right) \left( \partial_{a} \lambda(x) \, \delta^{l}{}_{ \vert d]} - \partial^{l} \lambda(x) \, g_{\vert d]a} \right)   \notag \\
\, & + \frac{\lambda^{-1}}{2} H^{b}{}_{ [c \vert l} \left( \partial_{a}\lambda(x) \, \delta^{l}_{\; \vert d]} - \partial^{l} \lambda(x) \, g_{\vert d] a}\right) + \frac{\lambda^{-1}}{2} \left( \partial_{l } \lambda(x) \, \delta^{b}{}_{ [c \vert } - \partial^{b} \lambda(x) \, g_{l [c \vert} \right) H^{l}{}_{\vert d] a} . \label{Riem_gB}
\end{align}
The Ricci tensor $\text{Ric}_{ad}$ corresponds to:
\begin{align}
\text{Ric}_{ad} = & \left( \text{Ric}^{\text{\tiny{L.C.}}} \right)_{ad} + \frac{1}{2} \nabla^{\text{\tiny{L.C.}}}_{l} H^{l}{}_{ da} - \frac{1}{4} H^{m}{}_{ d l} H^{l}{}_{ m a} +  (2-d) \nabla^{\text{\tiny{L.C.}}}_{d} \left(\lambda^{-1} \nabla^{\text{\tiny{L.C.}}}_{a} \lambda(x)\right)  \notag \\
\, & - \left(\nabla^{\text{\tiny{L.C.}}} \right)^{c} \left( \lambda^{-1} \nabla^{\text{\tiny{L.C.}}}_{c}  \lambda(x) \right)\, g_{ad} + \lambda^{-2} (2-d) \left( \left( \partial \lambda(x) \right)^{2} g_{ad} - \partial_{a} \lambda(x) \, \partial_{d} \lambda(x) \right) \notag \\
\, & - \lambda^{-1} \frac{4- d}{2} H^{l}{}_{ da} \partial_{l} \lambda(x) . \label{Ric_gB}
\end{align}

To briefly recap, in this subsection \ref{CONN_gBphi} we followed the prescriptions developed in section \ref{middle} and subsection \ref{Curvv} (choosing the first option listed there) and presented the explicit components expressions, in the case of a graded Poisson algebra deformed with a Riemannian metric $g$, a Kalb-Ramond field $B$ and a dilaton $\phi$, for the general connection $\widetilde{\nabla}$ \eqref{comps} and its restriction to vector fields \eqref{Nabla}, the Riemann tensor on vector fields $\text{R}^{b}_{\; a cd} $ \eqref{Riem_gB} and the Ricci tensor $\text{Ric}_{ad}$ \eqref{Ric_gB}. 

Two alternative but equivalent viewpoints on the connection for tangent vectors were  given. First, we worked out a Koszul formula from $[ \cdot, \cdot]^{\prime} - \llbracket \cdot, \cdot \rrbracket$, whose expression \eqref{KSZL}  resembles that of the standard Riemannian geometry case, but fixes all remaining ambiguities in the order of the entries (indices). Then we discussed torsion and metricity of the connection, justifying the statement that it is the metric connection w.r.t. $2 g(x)$ with contorsion tensor given by $ \text{K} = H + \frac{1}{2}\alpha$.

We are now nearly ready to show the main outcome of the deformation \eqref{br_GgB}. But before that, let us spend a few words on the low energy effective action for $10$-dimensional closed strings.

\subsection{Supergravity bosonic NS-NS sector}
\label{last}
In this subsection we would like to show that the connection $\widetilde{\nabla}$, for the deformation chosen, is relevant in building the $10$-dimensional supergravity action. In fact, it already contains all the fields of the supergravity multiplet, the metric $g$, the Neveu-Schwarz field $H$ and the scalar field $\phi$ known as the dilaton. Prior to that, however, let us collect some information on the physical model.

One way to introduce the supergravity action functional is by asking for cancellation of the worldsheet conformal anomaly for the superstring. The requirement of vanishing $\beta$-functions yield field equations for $\phi(x)$, $g_{ab}(x)$ and $B_{ab}(x)$, which can in turn be obtained by varying the following action \cite{Callan:1985ia}
\begin{equation}
\mathcal{S} = \frac{1}{2 \kappa_{d}} \int_{M} \text{Vol}_{d} \, \left[ R^{\text{\tiny{L.C.}}} - \frac{1}{12} H^{2} + 4 \left(\nabla^{\text{\tiny{L.C.}}} \phi \right)^{2} \right] e^{-2\phi}  - \frac{1}{4} \sum_{n=1}^{d} \, F^{2}_{n}\, .
\label{SUGRA}
\end{equation}
For the sake of completeness, in \eqref{SUGRA} also the $n$-form field strengths of $F_{n} = d A_{n}$ are taken into account. From a worldsheet perspective these are the fields due to a condensate of fermionic $0$-mode excitations of the type II superstring with Ramond boundary conditions on D-branes, in short R-R fluxes. This action is the low energy string effective action for the supersymmetric string.

In $10$ dimensions this action, correctly completed with the corresponding fermionic superpartners, gives the supergravity theory for both type IIA superstrings, with $\mathcal{N}=(1,1)$ supersymmetry, and type IIB superstrings, with $\mathcal{N}=(2,0)$ supersymmetry. The factor of $e^{-2 \phi} $ is a loop expansion parameter, and the specific factor of $2$ comes from the integral of the worldsheet curvature, equal to the Euler character $\chi_{\text{E}}$ of the sphere $\text{S}^{2}$, $\chi_{\text{E}}(\text{S}^{2}) = 2$ \cite{Blumenhagen:2013fgp}.

Furthermore \eqref{SUGRA} can be formulated independently from string theory: In fact, it was also developed as a field theory with local supersymmetry combined with general relativity, i.e. supergravity. Seen from this perspective, the same lagrangian is then valid for every dimension, the coupling constants getting different values depending on the dimensions.

In either cases the geometry involved is complex geometry. The construction that we followed here is based on Generalized Geometry which is a combination of symplectic and complex geometry: they can coexist if the cotangent bundle is doubled with the tangent. However we gave the expression for the Riemann and the Ricci tensors as in usual Riemannian geometry for the tangent bundle, see \eqref{Riem_gB} and \eqref{Ric_gB}. In $10$ dimensions these are tensors for the general linear group $GL(10)$ only, not of the full $O(10,10)$ of course. However it must be stressed that our prescriptions to obtain the connection $\widetilde{\nabla}$ just for tangent space, rely heavily on the differential geometry of Generalized Geometry. 

To reproduce the Lagrangian in the action $\mathcal{S}$ the scalar built from $\text{Ric}$ in \eqref{Ric_gB} must contain the volume form of the (compact subregion of the) $10$-dimensional spacetime over which we will integrate: this form is naturally a scalar density for $GL(10)$. We then need to perform the contraction in the following way:
\begin{equation}
\lambda^{-4} \left( \frac{g^{-1}}{2} \left( g + B \right) \frac{g^{-1}}{2} \right)^{ad} \, \text{Ric}_{ad}.
\label{Trace}
\end{equation}
This prescription for the trace relies in the following observation: in the computations the antisymmetric metric $g\pm B$ always occurred as direct metric while the only inverse metric employed was $ \frac{\lambda^{-2}}{2} g^{-1}$. As it may not be obvious, let us remark that every $2g(X,Y) $-like term, as displayed in the Koszul formula \eqref{KSZL}, comes from the symmetrization of $(g+B)(X,Y)$.\\
One could also possibly see this trace as in a Cartan-Palatini formulation, where the connection $1$-form comes from the combination of a pure torsion connection $\nabla$. Then one vielbein contributes with a $(g-B)$ factor.

The explicit computation of \eqref{Trace} gives the following lagrangian: 
\begin{align}
\sqrt{- \vert g \vert} \lambda^{d} \mathcal{L}^{(d)} = & \, \sqrt{- \vert g \vert } \lambda^{(d-4)} \left( \text{R}^{\text{\tiny{L.C.}}} - \frac{1}{12} H^{2} \right) + (1-d)(6-d)  \lambda^{(d-6)}\left( \nabla^{\text{\tiny{L.C.}}} \lambda(x) \right)^{2}  \sqrt{- \vert g \vert} \\
\, & + \sqrt{- \vert g \vert} \, \nabla^{\text{\tiny{L.C.}}}_{k} \left( H_{li}^{\; \; \, k} B^{il} \lambda^{(d-4)} - 2(d-1) \left[\lambda^{(d-5)} \, \left(\nabla^{\text{\tiny{L.C.}}}\right)^{ k} \lambda(x)\right] \right) \, .
\label{d_action}
\end{align}
In the corresponding action
\[
\mathcal{S}^{(d)} = \int_{M} \, \text{Vol}_{d} \, \mathcal{L}^{(d)} \, ,
\]
the last parenthesis in \eqref{d_action} is a boundary term which, for suitable boundary conditions on $g(x)$, $B(x)$ and $\phi(x)$, can be dropped. By setting $d=10$ and \[\lambda = e^{-\phi(x)/3}\] in \eqref{d_action} we recover the closed bosonic superstrings effective action:
\[
\mathcal{S}^{\text{(10)}} = \int_{M} \, \text{Vol}_{10} \, \mathcal{L}^{(10)} = \int_{M} \text{d}^{10} x\, e^{-2 \phi} \,  \sqrt{- \vert g \vert} \left( \text{R}^{\text{\tiny{L.C.}}} - \frac{1}{12} H^{2} + 4 \left( \nabla^{\text{\tiny{L.C.}}} \phi \right)^{2} \right) \, .
\]

\section{Discussion and comments}
\label{DISC}

In this article we have investigated a graded geometry approach to the quest of finding a natural interpretation of supergravity and string effective actions as some kind of Einstein General Relativity in a Generalized Geometry setting involving the double ``generalized tangent" bundle $TM \oplus T^{*}M$.
In this framework, we have presented natural constructions for generalized connections (of the first and second kind) as well as the associated torsion and curvature tensors.

Our initial setting is the graded Poisson structure of $C^{\infty}(T^{*}[2]T[1]M)$ 
and we consider local deformations of it. 
The deformation is implemented by an invertible change of the degree-$1$ coordinates
expressed in terms of local vielbeins. 
The introduction of this deformed graded Poisson algebra is interesting from several points of view: 
Firstly, the construction can be seen as an applications of Moser's lemma for symplectic forms in a graded setting. Secondly, there is an underlying bundle gerbe structure, which can be interpreted as a higher gauge theory with a $2$-form potential and $3$-form field strengh, complemented by a metric and a scalar field. Thirdly, the deformed graded Poisson structure is ready for quantization yielding a kind of first quantized gravity (although here we do not further pursue this direction).
 
Then, motivated by \v{S}evera and Roytenberg's result, via a derived bracket construction we related the new graded Poisson algebra to Courant algebroids with a deformed Dorfman bracket on $E \cong TM \oplus T^{*}M$. 
The deformed graded Poisson structure already naturally involves a metric connection, however, this connection is of Weitzenb\"ock type.
Via the derived bracket construction, and the introduction of a generalized Lie bracket \eqref{L_br}, we can build, out of this connection, a bona fide connection with \emph{curvature} and, once projected to $TM$, with standard \emph{torsion} given by~$H$. From this connection, evaluated on tangent vector fields, we obtain Riemann curvature and Ricci tensors. Suitably contracted, the resulting scalar curvature yields an action that reproduces the NS-NS sector of supergravity in $10$ dimensions \eqref{SUGRA}.  Some partial results without dilaton (i.e. with $\phi=0$) are already contained in our previous work \cite{phdthesis} and~\cite{Jurco:2015ywk}.

In view of  \v{S}evera's classification of exact Courant algebroids and Roytenberg's analysis of the correspondence of graded symplectic $NQ$-manifolds to Courant algebroids, one may wonder how it is possible to extract any geometric structures beyond the cohomology class $H^{3}(M, \mathbb{R})$ out of the deformed graded Poisson structure and Hamiltonian: The graded Poisson manifold and Hamiltonian that we study are after all directly related to an exact Courant algebroid and all exact Courant algebroids are isomorphic to the canonical $H$-deformed one up to $B$-transforms (see section \ref{very_first}). In other words the deformation can be undone by a change of graded coordinates: since we are using local coordinate changes, even $H$ can be transformed to zero. This is not very surprising: It is simply the inverse of the local change of coordinates that are used to deform the graded Poisson structure in the first place and consistency is ensured by a graded version of Moser's lemma. Note however that an inverse transformation that removes the deformation of the graded Poisson structure, Hamiltonian and Courant algebroid will deform the auxiliary geometric structure. The final results are coordinate-choice independent.

Some additional or auxiliary geometric structure is playing a role. A deformation becomes visible when compared to something undeformed. In the example of electromagnetism in the introduction the Poisson structure is deformed and the Hamiltonian is undeformed. Together, this yields the correct electromagnetic interaction and the additional structure is the Hamiltonian. In the main construction the necessary additional structures are also essentially canonical (undeformed up to dilaton rescaling). Here is a complete list of the additional structures beyond the deformed graded Poisson structure: A degree 3 Hamiltonian that satisfies the master equation (we chose the canonical one up to dilaton rescaling), a generalized Lie bracket (canonical with respect to the Darboux chart) of which we eventually only use the ordinary Lie bracket part, an embedding of $TM$ into the generalized tangent bundle $E$ compatible with the anchor map (we chose a non-isotropic splitting $s$ that is just the canonical projection up to rescaling by the dilaton) and finally a suitable trace prescription to obtain the Ricci scalar from the tensor (involves the non-symmetric metric $g+B$). Up to dilaton rescaling all these auxiliary geometric structures are canonical and undeformed.  It is probably also possible to use a construction, where the dilaton rescalings do not appear in the auxiliary geometric structures, but it will quite likely be more involved.

In this article we have focused on deformations that keep the anchor map undeformed up to rescalings. Another possibility that shall be investigated in a forthcoming article is a deformation based on the local vielbein
\[
\text{E} = \begin{pmatrix}
\mathbbm{1} & -\left(g+B \right)^{-1} \\
g-B & \mathbbm{1}
\end{pmatrix},
\]
whose corresponding metric $G$ is diagonal. This feature does not undermine the validity of the master equation as long as the anchor is more elaborate, namely it acts on a generalized vector $U = X + \sigma$ as
\[
\rho(U) = X - \left(g+B\right)^{-1}(\sigma).
\]
With this choice the computation of the connection and therefore the curvature is streamlined and dualization is immediate, but the expression for the anchored vector fields is now more involved. Notice also that the structure group is reduced to be $O(d) \times O(d)$, as usually in the literature on the topic of geometric frames for supergravity from Generalized Geometry/DFT.   

In fact, although supergravity was already obtained in the framework of Generalized Geometry as the analogue of Einstein's gravity, for the type II strings and the heterotic string (see, amongst the others, \cite{Jurco:2017gii}, \cite{Garcia-Fernandez:2013gja}, \cite{Grana:2008yw} and \cite{Severa:2018pag}, and for a double field theory formulation \cite{Deser:2016qkw}, \cite{Deser:2017fko}), here we present a different approach which begins with, and exploits, the $2$-graded symplectic $T^{*}[2]T[1]M$ with Hamiltonian $\Theta$ \eqref{HAAM}. Our construction is fully covariant, the metric and 2-form enter in the combination $g+B$ (as in string theory) and we do not impose any arguably arbitrary torsion and further constraints on a class of connections, but rather construct the connection directly from the deformation. This is also one of the striking differences with respect to \cite{Coimbra:2011nw}: here the metric $g$ appears already directly in the generalized vector basis, in a way that it is then present again in the generalized metric for $TM \oplus T^{*}M$. In other parts of that work and ours there can be found quite a few similarities, for example in the projection of the Ricci tensor, constructed in the former case for the metric generalized connection with no Gualtieri torsion and taking for arguments generalized vectors in the eigenbundles of the involution, while in the latter case it comes from the metric connection and is restricted to tangent vector fields, which (for $\phi =0$) have the same expression of the eigenvectors in the previous case. The relevance of the derived brackets in reproducing the action of infinitesimal symmetries on various configuration spaces of physical interest was pointed out recently, e.g. in \cite{Deser:2018oyg}. However, to our knowledge, so far there have been no attempts to construct those theories for which Generalized Geometry is accountable starting directly from the deformed graded Poisson algebra side. Since the derived bracket description from a corresponding deformation of the graded Poisson algebra is much less complicated than the techniques of differential geometry and cohomology on doubled bundles, we believe that theories of gravity (General Relativity, its modifications, Supergravity theories) could be very efficiently studied from this algebraic setup. As an example of the convenience of the method, let us recall again that metricity of the connection and Bianchi identities for the curvature are direct outcomes of the graded Jacobi identities.

The approach taken here keeps the covariance of the relevant objects for the differential geometry description, while at the same time suggesting slightly different formulas, e.g. the generalized Lie bracket \eqref{L_br} and the generalized torsion.
We find it also interesting that, for the particular ansatz chosen, several alternative original approaches of Einstein in his quest to find a theory of gravity are consistently combined. This fact deserves more attention. The combination $g+B$ plays the role of a non-symmetric metric, as in the nonsymmetric gravitational theory investigated by Einstein and others to find a unified field theory involving the electromagnetic field $F$ (which failed). We instead encounter the $3$-form $H$, as it should be for the abelian gerbe. 
Even ideas from teleparallelism show up via the Weitzenb\"{o}ck connection in the deformed Poisson structure.
Recall that in the teleparallel equivalent of gravity the source of the gravitational force is explained by means of the torsion rather than the curvature. (Note, however, that in this article we neither need nor assume a global vielbein and we certainly do not require a parallelizable manifold.) Here, all these rather different concepts and approaches appear in a mutually consistent manner and yield the bosonic part of the supergravity action.

The construction shares also some features with usual gauge theories, in the sense that the metric $g$, the $B$-field and the dilaton $\phi$ play a similar role as gauge potentials for the local symmetries of the dg-symplectic manifold. The relation is given by a graded generalization of Moser's lemma. The local vielbein $\text{E}(x)$ gathers all the gauge fields in a well-defined way. We shall investigate the resulting novel approach to gravity as a gauge theory elsewhere. 
We are convinced that more gravity theories with various kinds of geometry can be formulated starting from deformed graded Poisson algebras, and as such they will share the same unifying description. Some attempts in this direction can be found in \cite{Asakawa:2015jza} and \cite{Lust:2012fp}.
To conclude, let us mention some open questions:
A complete treatment of the R-R fields in our description is still missing. Once that this is done, one could attempt to explain also why the bosonic fields in the gauge description of $E$ derived from the deformed dg-symplectic $T^{*}[2]T[1]M$ appear to be in a supersymmetric (chiral or non-chiral) representation for $d=10$ SUGRA. Another task to be accomplished concerns the $C^{\alpha \beta \gamma} \, \xi_{\alpha} \xi_{\beta} \xi_{\gamma}$ fluxes: one could study them via the full Hamiltonian $\Theta$, or via the Poisson algebra. In some other works the authors have dealt with them in the Hamiltonian, e.g.\ \cite{Crow-Watamura:2018liw}, but the other scenario opens up more intriguing possibilities. It would also be interesting to know the dynamics of a test particle in $M$ with graded phase space $T^{*}[2]T[1]M$, 
i.e.\ to find its geodesic equation from Hamilton's equations. This may require the underlying membrane sigma model to be non-topological (at least off-shell), i.e.\ $\{ \Theta, \Theta \} \neq 0$.

\section{Appendix}
\label{app}
A collection of fully general formulas that appeared in the text is reproduced here. Everywhere, $U, V, W \in \Gamma(E)$, $f \in C^{\infty}(M)$, anchor $\rho : E \rightarrow TM$, bilinear pairing $\langle \cdot, \cdot \rangle : \Gamma(E) \times \Gamma(E) \rightarrow  C^{\infty}(M)$
\begin{itemize}
\item \emph{Connection}: 
\[
\nabla_{V} \left( f W \right) = (\rho(V) f) \, W + f \nabla_{V} W\, , \quad \nabla_{fV} W = f \nabla_{V} W 
\]
\[
 \  \text{ \footnotesize\emph{type 1:} } \ \Gamma(V;f W,U) = (\rho(V) f) \langle U, W\rangle \, , \
 \Gamma(V; W, fU) = \Gamma(f V; W,U) = f  \Gamma(V; W,U)  
\]
\item \emph{Metricity}:
\begin{align*}
\rho(V) \langle W, U \rangle &= \langle \nabla_V W, U \rangle + \langle W, \nabla_V U \rangle \\
&= \Gamma(V;W,U) + \Gamma(V;U,W)
\end{align*}
\item \emph{Generalized Dorfman bracket}:
\begin{align*}
[U,V] & = \langle V,\nabla U \rangle + \nabla_{U} V - \nabla_{V} U  \\
\text{i.e.: } \left\langle W, [U, V] \right\rangle &= \langle V,\nabla_{W} U \rangle + \langle W,\nabla_{U} V \rangle - \langle W,\nabla_{V} U \rangle 
\end{align*}
\item \emph{Generalized Lie bracket}:
\[
\llbracket U, V \rrbracket = - \llbracket V, U \rrbracket, \quad \llbracket U, f V \rrbracket = \left(\rho(U) f \right) V + f \llbracket U, V \rrbracket
\]
\[
\text{ \footnotesize\emph{in a holonomic basis (see section \ref{sec:gentorsion}):}} \; \;  \llbracket V,W \rrbracket = \rho(V) W - \rho(W) V   
\]
\item \emph{Relation between connection and brackets}:
\[
\tilde \Gamma(W;U,V) = \left\langle W, [U, V] \right\rangle - \langle W,\llbracket U,V \rrbracket \rangle 
\]
\item \emph{Torsion tensor}:
\[
\text{T}(V,W) = \nabla_{V} W - \nabla_{W} V - \llbracket V,W \rrbracket \, ,
\]
\item \emph{Curvature tensor}:
\[
R(V,W) = \nabla_{V} \nabla_{W} - \nabla_{W} \nabla_{V} - \nabla_{\llbracket V,W \rrbracket}\, .
\]
\end{itemize}

\acknowledgments 

E.B. and P.S. are grateful to the RTG 1620 ``Models of Gravity'' for support during the completion of this research project. We thank A. Chatzistavrakidis and M. Pinkwart for helpful discussions. This work is based in part on earlier work done in collaboration with F.-S. Khoo \cite{phdthesis}. 


\bibliographystyle{ieeetr}
\bibliography{Poiss_NS_JHEP-version}

\end{document}